\documentclass[letterpaper, journal, onecolumn, 11pt]{IEEEtran}
\usepackage{tikz}
\usepackage{amsmath}
\usepackage{outline}
\usepackage[utf8]{inputenc}
\usepackage{wrapfig}
\usepackage{amsmath,amssymb,amsbsy,mathrsfs,cases,amsthm,stmaryrd}
\usepackage{dsfont}
\usepackage{svg, hyperref}
\usepackage{graphicx}
\usepackage{multirow, multicol}
\usepackage{booktabs}
\usepackage{hhline}
\usepackage[normalem]{ulem}

\usepackage{filecontents}
\usepackage{listings}
\newcommand{\liu}[1]{{\color{black} #1}}
\newcommand{\bp}{ \begin{proof}}
	\newcommand{\ep}{\end{proof} }

\newcommand{\be}{\begin{equation}}
\newcommand{\ee}{\end{equation}}
\newcommand{\bqq}{\begin{eqnarray}}
\newcommand{\eqq}{\end{eqnarray}}
\newcommand{\bal}{\begin{align}}
\newcommand{\eal}{\end{align}}
\newcommand{\bqn}{\begin{eqnarray*}}
	\newcommand{\eqn}{\end{eqnarray*}}
\newcommand{\nn}{\nonumber}
\newcommand{\ba}{\left[ \begin{array}}
	\newcommand{\ea}{\\ \end{array} \right]}

\def\bxi        {{\boldsymbol \xi}}

\newcommand{\cE}{{\mathcal{E}}}

\newcommand{\cG}{{\mathcal{G}}}

\newcommand{\cM}{{\mathcal{M}}}
\newcommand{\cN}{{\mathcal{N}}}

\newcommand{\cV}{{\mathcal{V}}}

\newcommand{\grad}{{\nabla}}

\newcommand{\eq}[1]{\begin{align}#1\end{align}}
\newcommand{\beqn}{\begin{eqnarray}}
\newcommand{\eeqn}{\end{eqnarray}}

\newcommand{\RR}{\mathbb{R}}

\DeclareFontFamily{U}{mathx}{\hyphenchar\font45}
\DeclareFontShape{U}{mathx}{m}{n}{
	<5> <6> <7> <8> <9> <10>
	<10.95> <12> <14.4> <17.28> <20.74> <24.88>
	mathx10
}{}
\DeclareSymbolFont{mathx}{U}{mathx}{m}{n}
\DeclareFontSubstitution{U}{mathx}{m}{n}
\DeclareMathAccent{\widebar}{0}{mathx}{"73}

\def\Zint{{\mathchoice{\setbox1=\hbox{\sf Z}\copy1\kern-.75\wd1\box1}
		{\setbox1=\hbox{\sf Z}\copy1\kern-.75\wd1\box1}
		{\setbox1=\hbox{\scriptsize\sf Z}\copy1\kern-.75\wd1\box1}
		{\setbox1=\hbox{\scriptsize\sf Z}\copy1\kern-.75\wd1\box1}}}

\usepackage[utf8]{inputenc}

\usepackage{listings}
\usepackage{xcolor}

\definecolor{codegreen}{rgb}{0,0.6,0}
\definecolor{codegray}{rgb}{0.5,0.5,0.5}
\definecolor{codepurple}{rgb}{0.58,0,0.82}
\definecolor{backcolour}{rgb}{0.95,0.95,0.92}

\lstdefinestyle{mystyle}{
    backgroundcolor=\color{backcolour},   
    commentstyle=\color{codegreen},
    keywordstyle=\color{magenta},
    numberstyle=\tiny\color{codegray},
    stringstyle=\color{codepurple},
    basicstyle=\ttfamily\footnotesize,
    breakatwhitespace=false,         
    breaklines=true,                 
    captionpos=b,                    
    keepspaces=true,                 
    numbers=left,                    
    numbersep=5pt,                  
    showspaces=false,                
    showstringspaces=false,
    showtabs=false,                  
    tabsize=2
}

\usepackage{empheq}
\newcommand*\widefbox[1]{\fbox{\hspace{2em}#1\hspace{2em}}}

\begin{document}
%-------------------------------------------------------------------------------
\definecolor{codegreen}{rgb}{0,0.6,0}
\definecolor{codegray}{rgb}{0.5,0.5,0.5}
\definecolor{codepurple}{rgb}{0.58,0,0.82}
\definecolor{backcolour}{rgb}{0.95, 0.95, 0.95}
\definecolor{dkgreen}{rgb}{0,0.6,0}
\definecolor{gray}{rgb}{0.5,0.5,0.5}
\definecolor{mauve}{rgb}{0.58,0,0.82}

\lstset{frame=tb,
  backgroundcolor=\color{backcolour},
  language=python,
  aboveskip=3mm,
  belowskip=3mm,
  showstringspaces=false,
  columns=flexible,
  basicstyle={\footnotesize\ttfamily},
  numbers=none,
  numberstyle=\tiny\color{gray},
  keywordstyle=\color{blue},
%   identifierstyle=\color{blue},
  commentstyle=\color{dkgreen},
  stringstyle=\color{mauve},
  breaklines=true,
  breakatwhitespace=true,
  tabsize=2,
  captionpos=b
}

% %Code listing style named "mystyle"
% \lstdefinestyle{mystyle}{
%   backgroundcolor=\color{backcolour},   commentstyle=\color{codegreen},
%   keywordstyle=\color{magenta},
%   numberstyle=\tiny\color{codegray},
%   stringstyle=\color{codepurple},
%   basicstyle=\ttfamily\footnotesize,
%   breakatwhitespace=false,         
%   breaklines=true,                 
%   captionpos=b,                    
%   keepspaces=true,                 
%   numbers=left,                    
%   numbersep=5pt,                  
%   showspaces=false,                
%   showstringspaces=false,
%   showtabs=false,                  
%   tabsize=2
% }

% %"mystyle" code listing set
% \lstset{style=mystyle}

%don't want date printed

% make title bold and 14 pt font (Latex default is non-bold, 16 pt)
\title{BlueFog: Make Decentralized Algorithms Practical for
Optimization and Deep Learning}

\if@0
%for single author (just remove % characters)
\author{
{\rm Your N.\ Here}\\
Your Institution
\and
{\rm Second Name}\\
Second Institution
% copy the following lines to add more authors
% \and
% {\rm Name}\\
%Name Institution
} % end author
\fi

\author{Bicheng Ying$^{1,2} $, Kun Yuan$^{3}$,  Hanbin Hu$^{2,4}$, Yiming Chen$^{3}$, Wotao Yin$^3$\\[5mm]
{\small 
$^1$University of California, Los Angeles  $^2$Google Inc. \\
$^3$DAMO Academy, Alibaba Group $^4$University of California, Santa Barbara \\[1mm]
\texttt{ybc@ucla.edu, kun.yuan@alibaba-inc.com, hanbinhu@ucsb.edu, } \\ 
\texttt{$\{$charles.cym, wotao.yin$\}$@alibaba-inc.com}
}
}

\maketitle

%-------------------------------------------------------------------------------
\begin{abstract}
%-------------------------------------------------------------------------------\
\vspace{-5mm}
\noindent
{\color{black}
Decentralized algorithm is a form of computation that achieves a global goal through local dynamics that relies on low-cost communication between directly-connected agents. On large-scale  optimization  tasks  involving  distributed  datasets, decentralized algorithms have shown strong, sometimes superior, performance over distributed algorithms with a central node. Recently, developing decentralized algorithms for deep learning has attracted great attention. They are considered as low-communication-overhead alternatives to those using a parameter server or the Ring-Allreduce protocol.
However, the lack of an easy-to-use and efficient software package has kept most decentralized algorithms merely on paper.
%is nontrivial. In particular, some  recent  development  of decentralized algorithms  involves  asynchrony,  time-varying  network  topologies,  and asymmetric communication, introducing an even higher implementation barrier. 
%We fill the gap between the rapid development of algorithms in the research society and the lack of practice in real world applications with the introduction
To fill the gap, we introduce BlueFog, a python library for straightforward, high-performance implementations of diverse decentralized algorithms. 
%To make these advanced decentralized communication primitives relatively simple, we manage to design 
Based on a unified abstraction of various communication operations, BlueFog offers intuitive interfaces to implement a spectrum of decentralized algorithms, from those using a static, undirected graph for synchronous operations to those using dynamic and directed graphs for asynchronous operations.
%so that users can easily develop their own algorithms. 
BlueFog also adopts several system-level acceleration techniques to further optimize the performance on the deep learning tasks.
%by adopting several acceleration techniques such as computation/communication overlapping, tensor fusion, hierarchical communication, etc.
%In the  experimental study,  
On mainstream DNN training tasks, BlueFog  reaches  a  much  higher throughput and achieves an overall $1.2\times \sim 1.8\times$ speedup over Horovod, a state-of-the-art distributed deep learning package based on Ring-Allreduce.
BlueFog is open source at \url{https://github.com/Bluefog-Lib/bluefog}.
}

\end{abstract}

{
\section{Introduction}
Decentralized computational methods are distributed computational methods running without a centralized server. They do not directly perform global operations such as computing the average of distributed numbers. They can, however, obtain the same results through local dynamics, namely, a series of computation and agent-to-agent direct communication steps.
On large-scale optimization tasks involving distributed datasets, recent decentralized computational methods have shown strong, sometimes superior, performance~\cite{lian2017can,lian2018asynchronous,assran2019stochastic,chen2021accelerating,yuan2021decentlam}.

Modern deep-learning models are growing quickly in size \cite{devlin2019bert,NEURIPS2020_1457c0d6}, and training them require extremely large datasets \cite{deng2009imagenet, lahiri:2014:SRW, mundhenk2016large}.
With the rapid increase of our ability to gather data, some classic signal processing and statistical learning problems have also grown very large. However, the performance of each computing core stopped improving over a decade ago.
Therefore, solving large optimization problems in reasonable time requires parallel, distributed computation.
Most existing distributed optimization methods are based on iterative local computation by individual agents and global communication, for example, computing the average of distributed model parameters or taking the sum of distributed gradients. 

%As all agents are connected by the network, they can communicate one another directly or through the relays of other agents, so we can execute any algorithm by implementing communication between arbitrary agents. However,
Communication, rather than computation, tends to be the bottleneck. Many-to-one communication, one-to-many communication, and many rounds of communication (of even short messages) all incur huge costs.
The two common types of methods for computing global averages are Parameter Server~\cite{smola2010architecture} and Ring-Allreduce~\cite{ring-allreduce}. The former uses a central server and performs many-to-one and one-to-many communication. The latter places $n$ agents on a ring and uses $2n$ rounds of communication. Their running times grow linearly with the number of agents $n$. Calling them at every iteration of an iterative algorithm is very expensive.

Decentralized algorithms rely on inexpensive communication between directly-connected agents, and many different pairs of those agents communicate simultaneously. If every agent is connected only to a few others (in a so-called \emph{sparse} graph), it takes constant time for all directly-connected agents to communicate at least once. 

Decentralized algorithms can run on any connected network, even if the network topology changes dynamically. When a few agents or links are down, the algorithms run correctly on the remaining agents without modifications. Therefore, decentralized algorithms are more robust than those relying on a critical central agent or links.

 \begin{figure}[t]
 	\centering
 	\includegraphics[width=0.95\textwidth]{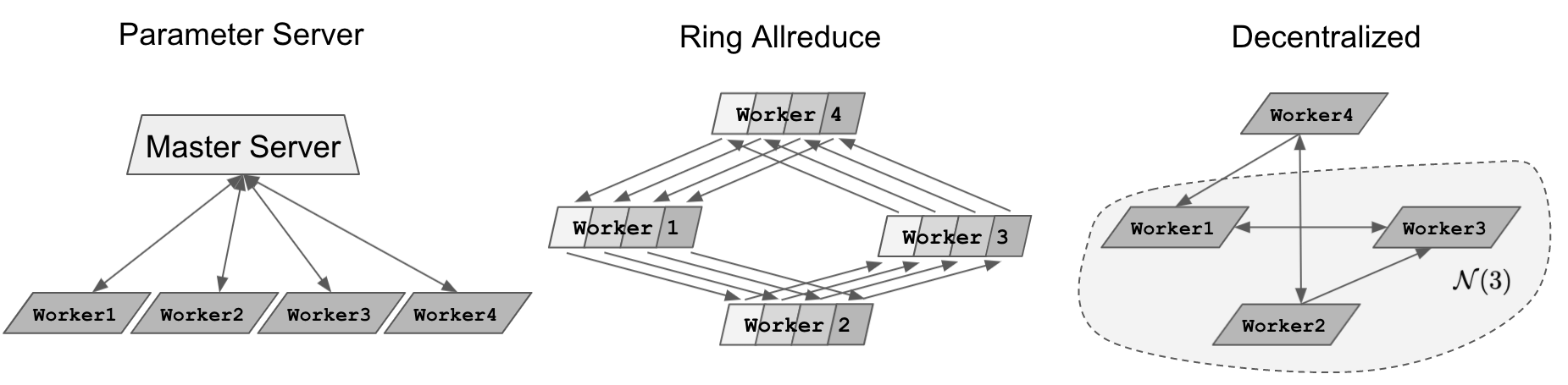}
 	\vspace{-2mm}
 	\caption{\small {Illustration of Parameter Server, Ring-Allreduce, and decentralized partial averaging. In partial averaging, nodes exchange information over edges; nodes do not relay information. As an example, {Node  $3$} collects information from its neighbors $1$ and $2$. Other nodes do the same but not depicted.}}
 	\label{fig:decen_comm} \vspace{-2mm}
 	\vspace{-2mm}
 \end{figure}

It it worth mentioning there are natural networks (e.g., fish school) and man-created networks (e.g., transportation network), where controlling all the agents by a central leader is practically infeasible or prohibitively expensive~\cite{acemoglu2011bayesian, kleinberg2006complex, inalhan2002decentralized, lu2011centralized}. Decentralized algorithms are, therefore, fundamental methods for sensing, signal processing, and controlling in those networks.

We can write most known decentralized algorithms in a few lines of equations. However, realizing them in a computer network is quite complicated. In particular, implementing a set of coordinated communication steps between specific pairs of nodes requires a skillful application of low-level communication libraries. Therefore, despite the development of decentralized algorithms over the past two decades, we are yet to see a software package for their rapid implementations. Most decentralized algorithm developers have been simulating their algorithms on a single computer in Matlab or Python. The very few packages that can run in a computer network were written for specific algorithms and network topologies. Adapting them to another project requires modifying lower-level communication. The recent development further involves asynchrony, time-varying network topologies, and asymmetric communication have created an even higher implementation barrier.

We fill the gap between the rapid development of algorithms on paper and the lack of practice in real world with the introduction of BlueFog, a python library for straightforward, high-performance implementations of diverse decentralized algorithms.
\begin{enumerate}
    \item Users with no experience with distributed systems can implement a decentralized algorithm in just a few clean lines of Python codes of a single program. The program will be run on multiple agents (with their different data) to achieve parallelism, which is commonly known as single program, multiple data (SPMD). 
    
    \item We design BlueFog to support a series of features for advanced users and future development. They include: the controls of partial-averaging coefficients, one-way communication in the pull and push forms, asynchrony, time-varying networks.
    % , and the overlapping of communication and computation. 
    We managed to keep the advanced communication primitives relatively simple through a unified abstraction of different decentralized communication operations.
    
    \item BlueFog is designed to be used together with modern deep learning framework like PyTroch to train large deep neural networks. All features of these framework (e.g., network design, auto-differentiation, various optimizers) are unaffected. % By changing a few lines of codes, the user can switch among parameter server, ring-allreduce, and decentralized methods. 
    
    \item We optimized BlueFog for efficient communication. BlueFog can be connected to a Message Passing Interface (MPI) library \cite{gabriel04:_open_mpi} for multi-CPU computing or the NVIDIA Collective Communications Library (NCCL) \cite{nccl27} for multi-GPU computing. On top of that, we implemented computation/communication overlapping, tensor fusion for multiple small tensors, hierarchical communication, and other acceleration techniques in BlueFog.
\end{enumerate}
The BlueFog package comes with several examples as a tutorial. They include major decentralized algorithms for signal processing: decentralized gradient descent \cite{nedic2009distributed,chen2012diffusion}, exact diffusion \cite{yuan2018exact,li2019decentralized}, and gradient tracking \cite{xu2015augmented,di2016next,nedic2017achieving,qu2018harnessing,lu2020decentralized}. We also include a deep-learning-based image classification example where BlueFog is $1.2\sim 1.8\times$ faster than Horovod~\cite{sergeev2018horovod}, a state-of-the-art distributed training package based on Ring-Allreduce.

We comment on speed comparisons. Different distributed algorithms may return solutions in various manners. Those using a central agent typically form the solution at the central agent. In decentralized algorithms, all the agents converge to (nearly) the same solution uniformly; we take the solution at the rank-0 node. Comparing the performances of these two types of algorithms is not merely a task of comparing their costs-per-iteration (though we do in this paper) but also the numbers of iterations or total times for reaching a solution of the same accuracy. The theoretical analyses of iteration complexities and total running times are out of the scope of this paper; see~\cite{lian2017can,assran2019stochastic,koloskova2020unified,kong2021consensus,chen2021accelerating,yuan2021decentlam,alghunaim2021unified} for some recent decentralized algorithms that are proven faster than distributed algorithms. This paper, however, provides some numerical comparison results that demonstrate BlueFog.

{\color{black}BlueFog is an open-source project that was first available on GitHub in December 2019. It keeps evolving during the past two years, and the progress was reported in some conferences\footnote{the 18th China symposium on Machine Learning and Application (2020) and the East Coast Optimization Meeting (2021)}. BlueFog has provided demos for several distributed and decentralized algorithms introduced in \cite{ryu2021large}, and it has supported all deep learning experiments in \cite{chen2021accelerating,yuan2021decentlam,ying2021exponential}. All BlueFog code and supporting documents can be found at \url{https://github.com/Bluefog-Lib}.
}

\subsection{Organization}
%This paper targets to briefly introduce BlueFog and clarify how it is designed and implemented.  
The rest of paper is organized as follows. Sec.~\ref{sec:prelim-related-work} reviews decentralized optimization and algorithms. Sec.~\ref{sec:decen_comm_abs} provides a unified abstraction of the partial averaging operation that supports static or time-varying, push-style and pull-style, synchronous and asynchronous decentralized communication modes. Sec.~\ref{sec-application} illustrates the usage of BlueFog partial averaging with various applications in optimizations and signal processing. Sec.~\ref{sec:system-design} highlights BlueFog's system design that wraps the low-level decentralized communication primitives for providing state-of-the-art training performance for large-scale deep learning. Sec.~\ref{sec:implementation} discusses  implementation details of BlueFog. The performance of BlueFog in deep learning training tasks are provided in Sec.~\ref{sec-evaluation}. We conclude the paper in Sec.~\ref{sec-conclusion}.

\section{A Brief Review on Decentralized Optimization}
\label{sec:prelim-related-work}
%Before we present the BlueFog system design, we give a quick review of the decentralized optimization.  Here we use the decentralized gradient descent (DGD) as example to illustrate most relevant concepts to consider for the decentralized algorithms.
\subsection{Concepts and Theoretical Foundations}\label{sec:review-concept}
	
	{This section gives a quick review of the components of decentralized optimization through the example %. While BlueFog can support various existing decentralized algorithms, we utilize
	of the decentralized (stochastic) gradient descent.% as an example to illustrate relevant concepts and theoretical results.
	}
	
	\vspace{1mm}
	\noindent \textbf{Problem.}
	Partition a set of data $D$ to $n$ computing nodes, where node $i$ has access to local data $D_i$, $i=1,\dots,n$.
	Suppose they collaborate to solve the distributed optimization problem:
	\begin{align}\label{dist-opt}
	{\color{black} \min_{x \in \mathbb{R}^d}\  f(x)=\frac{1}{n}\sum_{i=1}^n f_i(x) \quad \mbox{where} \quad f_i(x): = \frac{1}{|D_i|}\sum_{\xi_i \in D_i} F(x;\xi_i).}
	\end{align}
	%Function $f_i(x)$ is local to node $i$, and random variable $\xi_i$ denotes the local data that follows distribution $D_i$. Each local distribution $D_i$ can be different from each other across all nodes. 
	Node $i$ can evaluate stochastic gradient $\nabla F(x;\xi_i)$, by sampling $\xi_i$ randomly, or compute the real gradient $\nabla f_i(x)$.
	We use stochastic gradient in the rest of this section.%; it must communicate to access information from other nodes.
	
	\vspace{1mm}
	{\noindent \textbf{\color{black}Distributed method based on global averaging.}} Each agent independently computes $g_i^{(k)} = \nabla F(x_i^{(k)};\xi_i^{(k)})$, and they will synchronize across the entire network to achieve the globally averaged gradient to update $x$. This method can be described by the equations:
% 	send their $g_i^{(k)}$ to a central agent called parameter server; the parameter server computes their average and updates $x$ before it is broadcast to all the agents. This method can be described by the equations:
\begin{align}
	g_i^{(k)} &= \nabla F(x_i^{(k)}; \xi_i^{(k)}),\quad i=1,\dots,n && \mbox{(local stochastic gradient)} \label{sgd-1}\\
	x^{(k+1)} &= x^{(k)} - \frac{\gamma}{n}\sum_{i=1}^n g_i^{(k)} && \mbox{(global averaging)} \label{sgd-2}
\end{align}
where $\gamma$ is the stepsize or learning rate. {\color{black}The global averaging can be implemented via Parameter Server or Ring-Allreduce, see Fig.~\ref{fig:decen_comm} for illustrations. Global averaging typically incurs significant communication overhead, which motivates the following decentralized approach.} 
% It is possible to compute the average without a central agent by methods such as ring-allreduce and tree-allreduce. 

	\vspace{1mm}
	\noindent \textbf{Decentralized method based on partial averaging.} In decentralized stochastic gradient descent \cite{lopes2008diffusion,chen2012diffusion,nedic2009distributed}, each agent computes a stochastic gradient, updates its local copy of $x$, and performs a partial averaging that involves the copies from its direct neighbors and itself. We can describe these steps by:
% 	is among the most widely-used decentralized methods. Given weights $\{w_{ij}\}$ and the set of neighbors $\mathcal{N}_i$, each node $i$ in decentralized stochastic gradient descent  will iterate in parallel as follows: 
	\begin{align}
	x_i^{(k+\frac{1}{2})} &= x_i^{(k)} - \gamma \nabla F(x_i^{(k)}; \xi_i^{(k)}),\quad i=1,\dots,n && \mbox{(local update)} \label{dsgd-1}\\
	x_i^{(k+1)} &= w_{ii} x_i^{(k+\frac{1}{2})} + \sum_{j\in \mathcal{N}(i)} w_{ij}\, x_j^{(k+\frac{1}{2})},\quad i=1,\dots,n && \mbox{(partial averaging)} \label{dsgd-2}
	\end{align}
	where $\mathcal{N}(i)$ is the set of neighbor agents of agent $i$ and the weights $w_{ij}$ are described below.
	The network topology and weights significantly affect the convergence performance and communication efficiency. 
	%$x_i^{(k)}$ is the local model of node $i$ at iteration $k$, $\xi_i^{(k)}$ is the realization of $\xi_i$ at iteration $k$, and $\gamma$ is the learning rate. When the full gradient $\nabla f_i(x_i^{(k)})$ can be accessed, decentralized stochastic gradient descent can be reduced to decentralized gradient descent \cite{yuan2016convergence}. 
		
	\begin{figure}[t]
		\centering
		\includegraphics[width=0.9\textwidth]{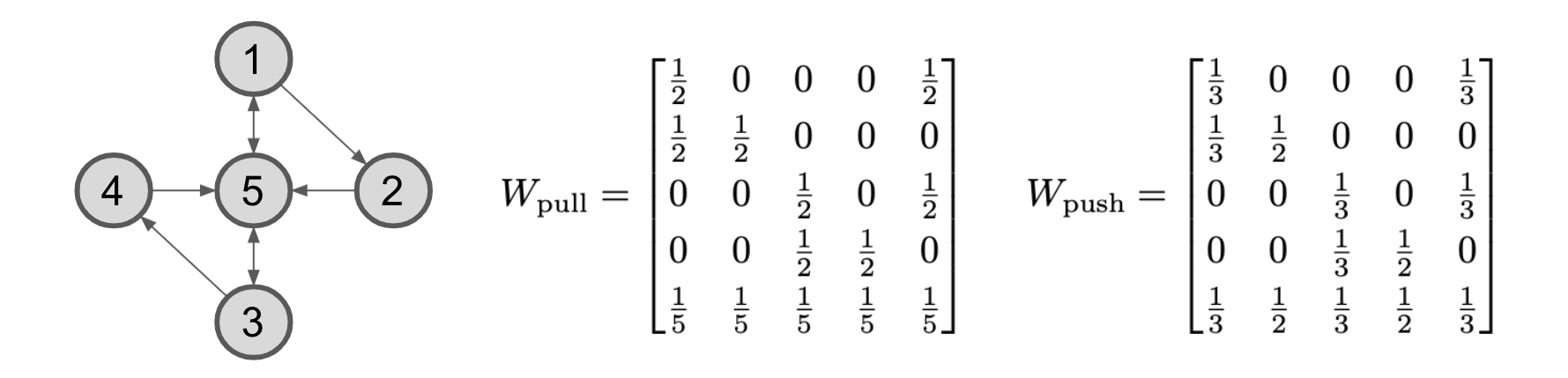}
		\vspace{-2mm}
		\caption{\small {An example topology with 5 nodes and two possible associated weighted matrices (pull weight matrix $W_{\rm pull}$ in the middle and push weight matrix $W_{\rm push}$ in the right). Note the topology is directed and $W_{\rm push}$ are not the transpose of $W_{\rm pull}$. }}
		\label{fig:topo_and_weights}
		\vspace{-2mm}
	\end{figure}
	
	\vspace{1mm}
	{\noindent \textbf{Network topology.} %Decentralized methods are based on partial averaging within neighborhood that is defined by the network topology. The topology can be of any shape, but its degree and connectivity will affect the communication efficiency and convergence rate of the decentralized algorithm. 
	Consider a directed graph $\cG = (\cV, \cE)$, where $\cV$ is the set of nodes and $\cE$ is the set of directed edges. An edge $(i,j) \in \cE$ implies that node $i$ can send information to node $j$. For each node $i$, define
	%its two sets of neighbors as follows: 
	\begin{align}
		\cN(i) & = \{j: (j,i) \in \cE\}\;\;\;\; \mbox{as its set of in-coming neighbors}, \label{in-neighbor}\\
		\cM(i) & = \{j: (i,j) \in \cE\}\;\;\;\; \mbox{as its set of out-going neighbors.} \label{out-neighbor}
	\end{align}
	Take node $5$ as an example in Fig.~\ref{fig:topo_and_weights}. It holds that $\cN(5) = \{1,2,3,4\}$ and $\cM(5) = \{1,3\}$. 
	
	\vspace{1mm}
	\noindent \textbf{Weight matrix.} Given $\cG$, define %$w_{ij}\in \mathbb{R}$ obeying
	%, the weight to scale information flowing from node $j$ to node $i$, as follows:
	%\vspace{-1mm}
	\begin{align}\label{wij}
	w_{ij}
	\begin{cases}
	\in \mathbb{R} & \mbox{if $(j,i) \in \cE$, or $i=j$} \\
	= 0 & \mbox{otherwise.}
	\end{cases}
	\end{align}
	Weight $w_{ij}$ is applied to the copy of $x_j$ sent from node $j$ to node $i$ (notice the positions of $j$ and $i$ in the subscript); the weight is zero if node $j$ cannot send messages to node $i$.
	All weights form the matrix $W := [w_{ij}]_{i,j=1} \in \mathbb{R}^{n\times n}$.
	%to stack all weights into a matrix. 
	%Such matrix $W$ will characterize the sparsity and connectivity of the underlying network topology.
	Commonly-used weight matrix $W$ falls into one of the following three categories: 
	\begin{enumerate}
		\item \textbf{Pull matrix} $W$ satisfies $W \mathds{1}= \mathds{1}$, i.e., every row adds up to $1$. Pull matrix $W$ is used with a directed graph. 
		
		\vspace{1mm}
		\item \textbf{Push matrix} $W$ satisfies $\mathds{1}^T W = \mathds{1}^T$, i.e., each column of a push matrix  $W$ adds up to $1$. Push matrix $W$ is also used with directed graph. 
%		\begin{align}\label{push-W}
%			\sum_{i=1}^n w_{ij} \overset{\eqref{wij-0}}{=} w_{jj} + \sum_{i\in \cM(j)}w_{ij} =1 \mbox{ for any node $j\in \cV$}.
%		\end{align}
%		In other words, each column of $W$ adds up to $1$. Inspired by \eqref{push-W}, a column-stochastic weight matrix can be generated as follows: a sender node $j$ will determine the weight for itself (i.e., $w_{jj}$) and edges connecting to its out-going neighbors (i.e., $w_{ij}$ for $i\in \cM(j)$) and guarantees \eqref{push-W} to be valid. 
		
		\vspace{1mm} 
		\item \textbf{Standard weight matrix} $W$ satisfies both $W \mathds{1}= \mathds{1}$ and $\mathds{1}^T W = \mathds{1}^T$ and used for undirected graph, as well as special directed graphs such as the  exponential graph \cite{ying2021exponential}.
	\end{enumerate}
	See   Fig.~\ref{fig:topo_and_weights} for examples of pull and push matrices. 
    When $W\ge 0$, a pull matrix is also known as a row-stochastic matrix, a push matrix as a column-stochastic matrix, and a standard weight matrix as a doubly-stochastic matrix. There are lots of flexibility into the selection of $W$ to serve different needs. %The three families of weight matrices will enable decentralized methods flexible enough to solve optimization problems over most graphs. 
    
    Given $W$, we can also deduce a directed graph $\cG = (\cV, \cE)$ with $\cV = \{1,\dots,n\}$ and $\cE = \{(j,i):w_{ij}\neq 0\}$. 
	}
	
%	\newpage
%	\vspace{1mm}
%	\noindent \textbf{Network topology and weights.} Decentralized methods are based on partial averaging within neighborhood that is defined by the network topology. The  topology can be of any shape, but its degree and connectivity will affect the communication efficiency and convergence rate of the decentralized algorithm. For a given topology, we define $w_{ij} \in (0,1)$, the weight to scale information flowing from node $j$ to node $i$, as follows:
%	\vspace{-1mm}
%	\begin{align}\label{wij}
%	w_{ij}
%	\begin{cases}
%	> 0 & \mbox{if node $j$ is connected to $i$} \\
%	> 0 &  \mbox{if $i=j$} \\
%	= 0 & \mbox{otherwise.}\vspace{-2mm}
%	\end{cases}
%	\end{align}
%	We further define $\mathcal{N}_i:=\{j|w_{ij} > 0 \ \mbox{and}\ j\neq i\}$ as the set of neighbors of node $i$. We define  weight matrix $W := [w_{ij}]_{i,j=1}^{n} \in \mathbb{R}^{n\times n}$ to stack all weights into a matrix. Such matrix $W$ will characterize the sparsity and connectivity of the underlying network topology. An example of the topology and its associated weight matrix $W$ is illustrated in Fig.~\ref{fig:topo_and_weights}.
	
% 	\vspace{1mm}
% 	\noindent \textbf{Partial averaging.} Given weights $\{w_{ij}\}$ over the graph, the neighborhood partial averaging operation of node $i$ can be expressed as \vspace{-1mm}
% 	\begin{align}\label{partial-ave}
% 	% \hspace{-10mm}
% 	\mbox{Partial averaging:}\quad x_i^{+} \leftarrow w_{ii} x_i + \sum_{j\in  \mathcal{N}(i)}w_{ij} x_j.
% 	\end{align}
% 	Partial averaging has much lower communication overheads. 

	\vspace{1mm}
	\noindent \textbf{Communication efficiency.} 
	When the network topology is sparse (e.g., a ring or a one-peer exponential graph \cite{assran2019stochastic,ying2021exponential}), each partial averaging step \eqref{dsgd-2} incurs $O(1)$ latency and $O(1)$ transmission time (the inverse of bandwidth), which are independent of $n$. %Consequently, decentralized methods are more communication efficient than those based on global averaging; see the communication efficiency paragraph.
	%The communication occurs in the partial averaging step \eqref{dsgd-2} in decentralized (stochastic) gradient descent. 
	Since each node only synchronizes with its direct neighbors, there is low synchronization overhead. %compared to the collective primitives based on global averaging. 
	Table \ref{Tb: Table-communication-overhead} compares the communication times of partial averaging and global averaging; see more discussions in Sec.~\ref{sec-dec-opt-impl}. It is observed that all global averaging primitives scale at $O(n)$, suffering from either a bandwidth-bound long delay or a long latency when $n$ is large.
% 	in which $n$ is the number of workers, $M$ is the size of the message to send, $L$ is the latency between a pair of nodes, and $B$ is the bandwidth of the communication channels \cite{ben2019demystifying}. It is observed that all global averaging primitives scale at $O(n)$, suffering from either a bandwidth-bound long delay or a long latency when $n$ is large. $O(n)$ is unavoidable for %resulted from the computation of x
% 	computing the {\bf exact global average} of all local gradients across $n$ nodes. The partial averaging primitive provided by BlueFog, on the other hand, achieves an $O(1)$ communication overhead over sparse topologies, which is independent of the topology size $n$. The superiority of partial averaging will be more evident when $n$ is large. 

	\begin{table}[h!]
	\centering
	\caption{\footnotesize Communication costs comparison \cite{ben2019demystifying}. $n$ is the number of nodes; $M$ is message size; $B$ is bandwidth; $L$ is latency of direct communication.}
	\begin{tabular}{rlcc}
	\toprule
	\textbf{Comm. primitive} & \textbf{Comm. cost} & \textbf{Averaging type} & \textbf{Note} \\ \midrule
	Parameter Server \cite{smola2010architecture}  & \quad\quad $\frac{nM}{B} + nL$ & global averaging & require a central node \vspace{1mm}\\
	Ring-Allreduce \liu{\cite{ring-allreduce}} & \quad\quad $\frac{2M}{B} + 2nL$ & global averaging &  \vspace{1mm} \\
	Byte-PS \cite{jiang2020unified} & \quad\quad  $\frac{M}{B} + nL$ & global averaging & require extra nodes \vspace{1mm}  \\ 
	{\color{black}BlueFog, partial averaging} & \quad\quad  {\color{black}$\frac{M}{B} + L$} & {\color{black}partial averaging} & \\ \bottomrule
	\end{tabular}\vspace{-2mm}
	\label{Tb: Table-communication-overhead}
	\end{table}

	\vspace{1mm}
	\noindent \textbf{Iteration complexity.} Iteration complexities, though unaffected by implementations, are essential properties of decentralized algorithms. We mention some recent results in passing. %The reduced communication in decentralized optimization and learning   
	% comes at a cost: slower convergence rate (see the rate expression in \cite[Theorem~2]{koloskova2020unified}). It is because the
	Although partial averaging alone is less effective in aggregating information than global averaging, %There is an extensive research on how to accelerate 
	some decentralized algorithms can match or exceed the performance of global-averaging-based distributed algorithms:  \cite{lian2017can,koloskova2020unified} established that decentralized SGD can achieve the same asymptotic linear speedup in convergence rate as (parameter server based) distributed SGD; % based on global averaging.
	\cite{assran2019stochastic, ying2021exponential} used exponential graph topologies to realize both efficient communication and effective aggregation by partial averaging; \cite{huang2021improving,yuan2021removing,alghunaim2021unified,vogels2021relaysum} improved the convergence rate of decentralized SGD by removing data heterogeneity between nodes; \cite{lu2021optimal,chen2021accelerating,kong2021consensus,xin2021stochastic} enhanced the effectiveness of partial averaging by periodically calling global averaging. BlueFog can implement all these algorithms including those use global averaging.
	%Note that \textbf{BlueFog is decoupled from these algorithms and strategies}. It only provides an efficient and unified implementation of the partial averaging operation. Users can utilize BlueFog to implement decentralized algorithms discussed above to achieve both efficient communication  and fast convergence. 
	
	\vspace{1mm}
	\noindent \textbf{Brief history of decentralized optimization.} Decentralized optimization can be traced back to \cite{tsitsiklis1986distributed}. Since then, it has been intensively studied in the control and signal processing communities. The first decentralized algorithms on general optimization problems include decentralized gradient descent \cite{nedic2009distributed}, diffusion \cite{lopes2008diffusion,chen2012diffusion,sayed2014adaptation}, and dual averaging \cite{duchi2011dual}. Various primal-dual algorithms come out to further speed up the convergence, and they are based on alternating direction method of multipliers (ADMM) \cite{mateos2010distributed,shi2014linear}, explicit bias-correction \cite{shi2015extra,yuan2017exact1,li2017decentralized}, gradient tracking \cite{xu2015augmented,di2016next,nedic2017achieving,qu2018harnessing}, and dual acceleration \cite{scaman2017optimal,uribe2020dual}. In deep learning tasks, decentralize SGD also attracted a lot of attentions recently. Many efforts have been made to extend the algorithm to time-varying topologies \cite{koloskova2020unified}, directed topologies \cite{assran2019stochastic,lu2020decentralized},  asynchronous settings \cite{lian2018asynchronous}, and data-heterogeneous scenarios \cite{tang2018d,xin2020improved,alghunaim2021unified,huang2021improving,yuan2021removing,lu2019gnsd,vogels2021relaysum}. 
% With careful consensus control  \cite{kong2021consensus} or periodic global averaging \cite{chen2021accelerating}, decentralize SGD can achieve $1.3\sim2\times$ training time speedup without severe performance degradation. 
Techniques such as quantization/compression \cite{alistarh2017qsgd,bernstein2018signsgd,koloskova2019decentralized2,koloskova2019decentralized,tang2019doublesqueeze}, periodic updates \cite{stich2019local,koloskova2020unified,yu2019linear}, and lazy communication \cite{chen2018lag,liu2019communication} were also integrated into decentralized SGD to improve communications. 

% There are also a few studies on accelerated variants of decentralized SGD, and most of them are on (static) momentum acceleration. \cite{assran2019stochastic,gao2020periodic} propose to run a local momentum SGD step first before the partial averaging is conducted. Another work \cite{yu2019linear} imposes an additional partial averaging over momentum to increase stability. Recent works \cite{lin2021quasi,yuan2021decentlam} developed strategies that can remove the momentum-incurred bias in decentralized momentum SGD. {All these methods are not with adaptive strategies to scale gradients.}
	
\subsection{Related works}
\label{sec-dec-opt-impl}
\vspace{1mm}
\noindent \textbf{Libraries that support global averaging.} Parameter Server (PS) \cite{smola2010architecture} is a well-known architecture adopted early in TensorFlow \cite{abadi2016tensorflow} where all workers communicate with the central parameter
server(s). It easily suffers from communication bottlenecks. If every worker sends a message of size $M$ to the central server and the network interface is saturated by every message, the total communication in PS is $n(\frac{M}{B} + L)$. %The PS architecture was adopted as one of the distributed training architectures in TensorFlow\cite{abadi2016tensorflow}. 
Like PS, the Ring-Allreduce \cite{ring-allreduce} architecture organizes all nodes on a ring and divides each local gradient tensor into $n$ chunks for parallel communication. %At each iteration, each local gradient chunk is passed over the ring and get accumulated at a unique node. After that, each accumulated gradient chunk will be broadcast to all the nodes. This algorithm 
Ring-Allreduce achieves a remarkable communication time of $\frac{2M}{B} + 2nL$. The bandwidth-bound time $\frac{2M}{B}$ is optimal \cite{patarasuk2009bandwidth}, where constant $2$ is not improvable (without hardware additions). Ring-Allreduce has been implemented in Horovod and Pytorch. Fig.~\ref{fig:decen_comm} illustrates the Parameter Server and Ring-Allreduce operations. 
Another framework {BytePS} \cite{jiang2020unified} is based on PS-lite  \cite{li2014scaling} and can reduce the bandwidth cost from $\frac{2M}{B}$ to $\frac{M}{B}$ using $n$ {\em additional} CPU servers. 
Instead of passing each of the $n$ chunks over a ring, each worker pushes its $i$th chunk to server $i$ and pulls the accumulated chunk back, 
achieving a total communication time of $\frac{M}{B} + nL$. The communication overheads of Parameter Server, Ring-Allreduce, and BytePS are listed in Table \ref{Tb: Table-communication-overhead}. 

\vspace{1mm}
\noindent \textbf{Libraries that support partial averaging.} The open-source codes or libraries to provide the system-level partial averaging are limited. The codes brought along with \cite{assran2019stochastic, lin2021quasi} implemented only the decentralized algorithms proposed therein. Prague \cite{luo2020prague} wraps {\em all-reduce} or {\em broadcast} operations, which is smart but relatively inefficient and restrictive. 
%The partial averaging implemented in these works is not flexible.  
%These works do not provide a unified abstraction of the partial averaging operation that can support static or time-varying topology, pull- or push-style communication, and synchronous or asynchronous modes.  
In particular, Prague forms a ring from a random subset of nodes and then ring-allreduces over them, thus ruling out other effective topologies such as the exponential graph \cite{assran2019stochastic,ying2021exponential}.  

BAGUA~\cite{gan2021bagua} is a recent open-source library that supports both global and partial averaging, offers full- and low-precision operations, and focuses on efficient deep learning. It does not support asynchronous communication, diverse and time-varying network topologies, and directed communications in pull- and push styles, which are supported by BlueFog to implement algorithms such as push-sum \cite{assran2019stochastic} and push-pull \cite{pu2020push,xin2018linear}, as well as more recent decentralized algorithms using those features. BlueFog supports classic decentralized optimization and control with a tutorial that covers several examples, in addition to deep learning.

%recent independent work  established an open-source library {BAGUA} to provide more efficient communication primitives than traditional Parameter-Server and Ring-Allreduce. These primitives include both global and partial averaging, either of which offers full precision and low precision options.  There are several fundamental differences between BAGUA and BlueFog. First, BlueFog provides asynchronous communication primitives while BAGUA does not. Second, BlueFog supports partial averaging over any static or time-varying, directed or undirected topology. Furthermore, BlueFog realizes flexible decentralized communications in both pull- and push-style; it supports decentralized strategies such as push-sum \cite{assran2019stochastic} and push-pull \cite{pu2020push,xin2018linear}. BAGUA, however,  has limitations on the topology choice; it mainly supports the ring and random topology. Third, BlueFog supports classical  optimization as well as deep learning; it has provided a detailed tutorial\footnote{Github address: \url{https://github.com/Bluefog-Lib/bluefog-tutorial}} on how to implement algorithms such as exact diffusion, gradient tracking, and their variance-reduced stochastic variants which are widely-used in control and signal processing. In contrast, {BAGUA is mainly developed and optimized for deep learning.} 
}

\vspace{1mm}
\subsection{Notation and Terminologies} 
BlueFog lies in the intersection between decentralized optimization and high-performance computation communities, so we use terms from both communities. 
Throughout the paper, we use \emph{process}, \emph{node}, and \emph{agent} interchangeably. Each node has a unique ID called \emph{rank}, a number starting from 0. The number of nodes is called the \emph{size} of the graph. A \emph{super node} refers to a physical machine, which may include one or more nodes inside. Most low-level communication primitives are named following the convention in MPI. We call the out-going and in-coming nodes as the  \emph{destination} and \emph{source} nodes of that communication, respectively.
%also call the out-going neighbors as the \emph{destination} nodes of the communication and in-coming neighbors as the \emph{source} nodes of communication. 
% The terminology \emph{push-} and \emph{pull-style} communications in the decentralized optimization community mean different from those in the high-performance computing community (e.g., \cite{hintjens2013zeromq}). {\color{red}In decentralized algorithm development, information will be scaled based on the sender (or receiver) in the push-style (or pull-style) communication, which is not related to the producer/consumer (also known as the push/pull mode in high-performance computing) communication pattern. 
Lastly, we use the terms  \emph{topology, graph} or \emph{network} interchangeably to indicate how all nodes are connected.

\section{Abstraction of Decentralized Communication} \label{sec:decen_comm_abs}
Decentralized algorithms are diverse. For BlueFog to provide a clean and consistent interface, we begin with summarizing the major dimensions that describe the variations in how decentralized algorithms communicate:
%One goal of BlueFog is to support flexible decentralized communication with a clean and consistent interface. However, unlike the Allreduce paradigm which has a clear and fixed definition, the  communication in decentralized algorithms has many variants. The conflict between the diversity of decentralized communications and the  consistent communication interface is a big challenge to develop BlueFog. To address it, we identify the major features of various decentralized communications as follows: % , which can be categorized into following three features:
\begin{enumerate}
    \item {\bf Communicate over static or time-varying topology.} The underlying topology in decentralized communication can remain static or keep changing over iterations.
    \item {\bf Communicate in a push- or pull-style.} The weight matrix associated with the topology can be push stochastic or pull stochastic; see Sec.~\ref{sec:review-concept} for the discussion on the weight matrix. 
    % Weight matrix of different style will determine how the weights are set up in decentralized communication. 
    \item {\bf Communicate in a synchronous or asynchronous mode.} Each node in decentralized communication can communicate and update information with or without synchronization with its neighbors. 
\end{enumerate}
%Note these features are not mutually exclusive; they can be combined in a combinatorial manner.
An algorithm may include one or multiple of these features.
For example, push-DIGing \cite{nedic2017achieving} uses synchronous communication in the push style over a time-varying topology. 
We develop primitives to support the above features, with more advanced primitives offer richer features and also require more inputs from the user.

The discussion in this section focuses on how we define and organized the programming interfaces, instead of their syntax, which we introduce in the next section with examples.

\subsection{Decentralized communication over static topology}
% \begin{minipage}{0.4\linewidth}
% % \begin{figure}
% %     \centering
%     \includegraphics[width=0.9\linewidth]{images/static_decentralize_comm.png}
% %     \caption{Caption}
% %     \label{fig:my_label}
% % \end{figure}
% \end{minipage}
We start with the primitives for the simplest case -- synchronous communication over a static graph: %. As discussed in Sec.~\ref{sec:prelim-related-work}, one of the key components in a decentralized algorithm is its associated underlying graph
\eq{\label{graph-global}
    \cG = (\cV, \cE) \;\;\;\;\mbox{(Global view)}
}
%call $\cG$ in \eqref{graph-global} a graph under the
By global view, we mean the information of the graph, including all the nodes, edges, and weights, is made available to all the nodes.
%because it (and its weighted matrix) contains the information of all the nodes and edges. 
%In the static topology scenario, the setting of the graph will not be changed once it is set up.
Once set up, $\cG$ remain unchanged. The BlueFog primitive for this setup is:
\eq{
    &\boxed{
        \mbox{set\_topology(graph\_object)} \to {\rm\; bool}
    }\nn
}
where ``graph\_object'' also contains the weight matrix $W$. BlueFog includes built-in implementation of commonly-used topologies including ring, grid, static exponential graph, and a few others. The user can create their own graphs with BlueFog. 

After a graph is set up, each node can access the weights for its partial averaging. BlueFog provides the following collective communication primitive for partial averaging \eqref{dsgd-2}: 
% This weight matrix is  accessible to all agents when the program determines to use some simple pre-defined topologies like ring, static exponential graph, etc. After the global topology is set, it embeds all information for the decentralized communication. Hence, we just require the following simple collective communication primitive:
\eq{
    \boxed{
        \mbox{neighbor\_allreduce(tensor, name)} \to {\rm\; tensor}
    }\nn
}
While \emph{partial averaging} is a math term, its implementation \emph{neighbor allreduce}
%As we stated in the terminology, we call  equation \eqref{dsgd-2} as partial averaging while its corresponding primitive as neighbor allreduce, which
follows the MPI naming convention, where {\it reduce} stands for reducing multiple tensors into one, {\it all} means all processes (nodes) are involved, and {\it neighbor} modifies ``all'' to include only the neighbors. Unlike {\ttfamily allreduce}, {\ttfamily neighbor\_allreduce} returns different tensor values for different nodes in general as they may have different neighbors and use different weights. It is clearly \emph{not a stateless function}. %  Furthermore, it is {\em not  a stateless} function since the behavior of this function is determined by the global topology setting information. 

\subsection{Push- and pull-style communication over time-varying topology}
The primitives of last subsection cannot be applied to create time-varying graphs. We use a new primitive for time-varying graphs, which provides each node with a local view (e.g., the weights associated with its neighbors).

%While {\ttfamily set\_topology} $+$ {\ttfamily neighbor\_allreduce} is intuitive and easy to use, it suffers from at least two limitations. First, it requires all nodes know the global topology  information, which is unnecessary since each node only communicates with its local neighbors. Second, it cannot support time-varying topology, i.e., the exploitation of different topologies in different iterations  \cite{sayed2014adaptation, ying2021exponential,nedic2017achieving}. Apparently, it is not practical to let all nodes achieve a global information of the time-varying topology per iteration. 

\begin{figure}[ht]
    \centering
    \includegraphics[width=0.56\linewidth]{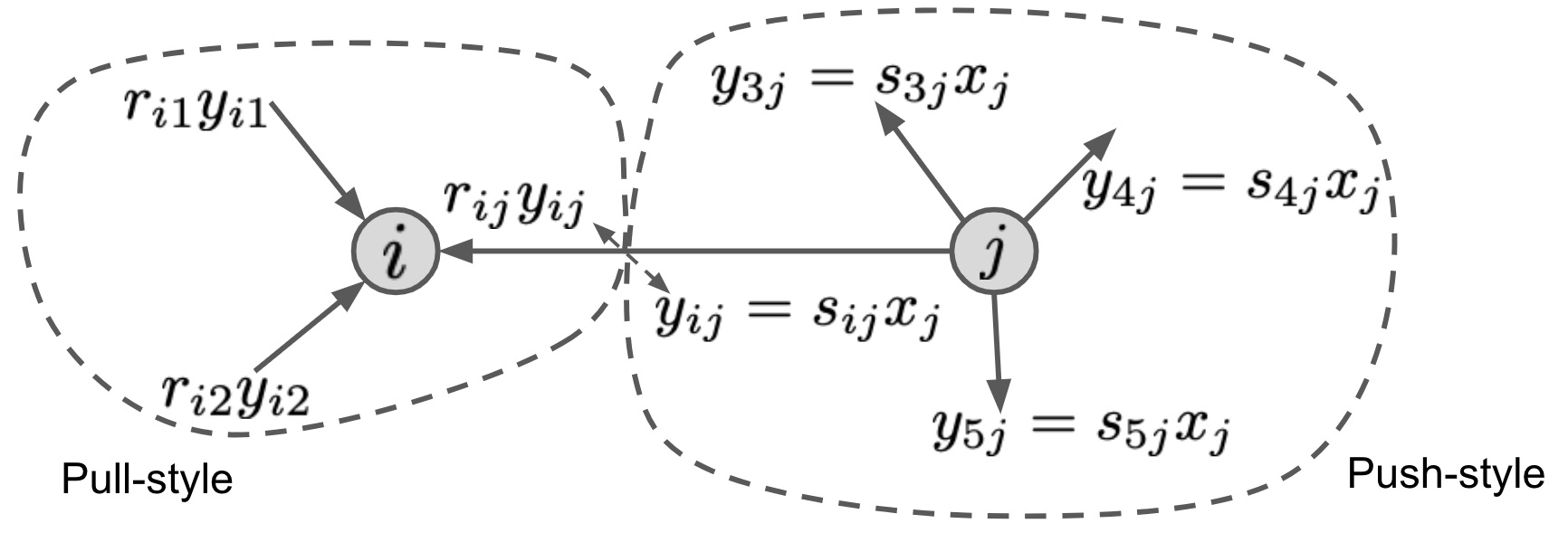}
    \caption{A illustration of the push- and pull-style communications. Node $i$ is pulling information from its in-coming neighbors while node $j$ is pushing information to its out-going neighbors. Note the pull-style circle and push-style circle are not necessary separable as plotted.}
    \label{fig:push-pull-pattern}
\end{figure}

Another way to define decentralized communication is through the local view of each process (node). Recall the set of in-coming neighbors and out-going neighbors introduced in \eqref{in-neighbor} and \eqref{out-neighbor} are all local information associated with each node $i$. Knowledge of these two sets is  sufficient to conduct partial averaging through local view. To this end, we extend partial averaging \eqref{dsgd-2} to the time-varying topology as follows
\eq{\label{2bzbz09}
    x_{i}^{(k+1)} = w^{(k)}_{ii} x_{i}^{(k)} + \sum_{j\in\cN(i)} r^{(k)}_{ij} s^{(k)}_{ij} x_{j}^{(k)}
}
where $r^{(k)}_{ij} \in [0, 1]$ and $s^{(k)}_{ij} \in [0, 1]$ both are the associated scaling weights when node $j$ sending the information to node $i$ at iteration $k$.
% %assigned to link $(j,i)$ (if it exists) %
% the receiving node $i$ and sending node $j$ at iteration $k$, respectively. 
Note that `$r$' stands for receiving-side scaling and `$s$' stands for sending-side scaling. If $r^{(k)}_{ij} s^{(k)}_{ij} = w_{ij}^{(k)}$ and $w_{ij}^{(k)}$ remains static for each iteration $k$, the recursion \eqref{2bzbz09} reduces to the partial averaging step in \eqref{dsgd-2}. By introducing a dummy variable $y_{ij}$, the above partial averaging operation is equivalent to 
\eq{
	y_{ij}^{(k)} =\;& s_{ij}^{(k)} x_j^{(k)},\;\; i \in \cM(j) \hspace{1.85cm} \mbox{(push-style communication)} \label{eq.dyn-push-comm} \\
    x_{i}^{(k+1)} =\;& w^{(k)}_{ii} x_{i}^{(k)} + \sum_{j\in\cN(i)} r_{ij}^{(k)} y_{ij}^{(k)} \hspace{1cm} \mbox{(pull-style communication)} \label{eq.dyn-pull-comm}
}
In the push-style communication \eqref{eq.dyn-push-comm}, a sending node $j$ will scale its own variable $x_j^{(k)}$ with weight $s^{(k)}_{ij}$ to achieve $y^{(k)}_{ij}$, and push it to each of its out-going neighbor. In the pull-style communication \eqref{eq.dyn-push-comm}, a receiving node $i$ will pull $y^{(k)}_{ij}$ from its in-coming neighbors and scales them with $r^{(k)}_{ij}$. Fig.~\ref{fig:push-pull-pattern} illustrates  the push- and pull-style communications. Two circles are drawn around the sending node $j$ and receiving node $i$, respectively. The left circle illustrates the pull-style communication \eqref{eq.dyn-push-comm} while the right illustrates the push-style communication \eqref{eq.dyn-pull-comm}. 
%We refer $r_{ij}$ as source weights with respect to node $i$ (`$r$' stands for receiving) and $s_{ij}$ as destination weights weights respect to node $j$ (`$s$' stands for sending).
%In a mathematical perspective, it doesn't extend the functionality since it is equivalent to \eqref{2oinvsd} if we simply set $w_{ij}  = r_{ij}s_{ij}$.  
%Nevertheless, splitting this scalar $w_{ij}$ into two parts is necessary for the library implementation in terms of the local view. To see the usefulness of this form, a simple illustration is given in Fig.~\ref{fig:push-pull-pattern}.

%because node $i$ pulls the information from the neighbors. Users can freely choose $r_{ij}$ to determine how to scale the received information and then combine. The right circle we call it push-style because node $j$ pushes its own information to its neighbors. It scales the information before reaching to the neighbors. 

The introduction of the dummy variable $y_{ij}$ completely decoupled the pull-style communication with the push-style communication. A pure push-style partial averaging can be achieved by setting $r^{(k)}_{ij} = 1$ and $s^{(k)}_{ij} = w^{(k)}_{ij}$. In addition, it is easy to guarantee $\sum_{i=1}^n s^{(k)}_{ij} = 1$, i.e., utilize push weight matrix, since the weights $\{s_{ij}\}_{i=1}^n$ is determined by the sending node $j$. Similarly, a pule pull-style partial averaging can be achieved by setting $r^{(k)}_{ij} = w^{(k)}_{ij}$ and $s^{(k)}_{ij} = 1$, and $\sum_{j=1}^n r^{(k)}_{ij} = 1$, i.e., utilization of the pull weight matrix, can be easily guaranteed since the weights $\{r_{ij}\}_{j=1}^n$ is determined by the receiving node $j$. When $r^{(j)}_{ij} \neq 1$ and $s^{(j)}_{ij} \neq 1$, both push- and pull-style communications exist in partial averaging. 

The {\ttfamily neighbor\_allreduce} primitive provided in BlueFog supports \eqref{eq.dyn-push-comm} and \eqref{eq.dyn-pull-comm} as follows
\eq{
    \boxed{
        \mbox{neighbor\_allreduce(tensor, name, [self\_weight, dst\_weights, src\_weights])} \to {\rm\; tensor}
    }\nn
}
%For the consistent primitive design, we use the same function name but extend it with three more optional argument. If none of these arguments is provided, the primitive will behave as we described in the static topology. If some of them are provided\footnote{It is only meaningful when three cases of the argument combination provided: 1) self\_weight and dst\_weights; 2) self\_weight and src\_weights; 3) self\_weight, src\_weights, and dst\_weights.}, {\ttfamily neighbor\_allreduce} will behave based on the provided argument and ingore the global topology setting. Here self\_weight is simply a scalar corresponding to $w_{ii}$ and both dst\_weights and src\_weights are the dictionary that mapping the neighbors' rank to the corresponding weights, i.e. $s_{ij}$ or $r_{ij}$
To be consistent with decentralized communication over static topology, we exploit the same function name but extend it with three more optional argument: {\ttfamily self\_weight}, {\ttfamily src\_weights}, and {\ttfamily dst\_weights}. Here {\ttfamily self\_weight} is simply a scalar corresponding to $w_{ii}$ for node $i$. In a local view of node $i$, {\ttfamily src\_weights} stands for the weights $\{r_{ij}\}_{j=1}^n$ to scale tensors received from in-coming source nodes; similarly, {\ttfamily dst\_weights} denotes $\{s_{ij}\}_{i=1}^n$ to scale tensors sent to out-going destination nodes. These three arguments enable partial averaging to be conducted with push-style, pull-style, or push-pull-style communication. In addition, the time-varying topology is now allowed since these three arguments can be passed to {\ttfamily neighbor\_allreduce} per iteration. Fourthermore, the {\ttfamily neighbor\_allreduce} primitive\footnote{Only four configurations of the arguments are meaningful: 1) no arguments for static topology usage; 2) self\_weight and dst\_weights for pure dynamic push-style communication; 3) self\_weight and src\_weights for pure dynamic pull-style communication; 4) self\_weight, src\_weights, and dst\_weights for dynamic push-pull-style communication.}
% {\color{red} Currently, the configuration of these parameters defines three different behaviours of the proposed} 
% \begin{enumerate}
% \item If none of these arguments is provided, the primitive will behave as we described in the static topology specified by the {\ttfamily set\_topology} primitive.
% \item If {\ttfamily self\_weight} and {\ttfamily src\_weights} are provided, it performs as a pure pull-style function assuming all $s_{ij}=1$. Here {\ttfamily src\_weights} provides the weights $r_{ij}$ and the corresponding topology for communication, which should be a sub-graph of the static original topology specified by the {\ttfamily set\_topology} primitive.
% \item If all the three parameters are provided, the communication topology is entirely deducted from the {\ttfamily src\_weights} and {\ttfamily dst\_weights}, and doesn't rely on the static topology set by the {\ttfamily set\_topology} primitive. Note that users need to ensure that the provided topology is valid, i.e., if process $i$ sends a tensor to process $j$, $r_{ij}$ should appear in {\ttfamily src\_weights} for process $j$, and $s_{ij}$ should appear in {\ttfamily dst\_weights} in process $i$. 
% \end{enumerate}
should also support automatic topology check (which is  discussed in details in Section \ref{sec.impl.nego}) to examine whether the weights specified by users are valid; otherwise the program may get stuck when the sending and receiving process do not match with each other.

\subsection{Asynchronous decentralized communication}
Asynchronous decentralized communication is much more complicated than the synchronous one \cite{lian2018asynchronous,williams2012c++,ben2019demystifying}. The key feature of asynchronous communication between processes is it decouples tensor movement with process synchronization. In  other words, one process (node) is allowed to move tensors without synchronization with remote processes (i.e., the neighboring nodes). 

This section proposes a solution inspired by the Remote Memory Access (RMA) technique \cite{mpi-3.0, gropp2014using}. To enable asynchronous decentralized communication, each node (process) first registers a {\it window}, which is a memory chunk created for neighbors. A node can create multiple windows,  and each window will be associated with a unique name and an immutable tensor. The total memory size of these windows is determined by the shape of the tensor as well as the number of in-coming neighbors. For example, if node 1 has two in-coming neighbors (say node 3 and node 5) and it owns a model parameter named ``convolution.layer1.weights'' with a shape of $5 \times 5 \times 3$, then it will register a continuous memory chunk capable to store a $150$-length vector\footnote{The exact memory size in bytes will be determined by the data type and systems.} ($150 = 5 \times 5 \times 3 \times 2$), with the first half storing model ``convolution.layer1.weights'' received from node 3 and the remaining half from node 5. The following two primitives are developed in BlueFog for the window creation and deletion.
\begin{subequations}
	\begin{empheq}[box=\widefbox]{align}
		\hspace{-7mm}\mbox{win\_create(tensor, name)} &\to {\rm\; bool} \hspace{-7mm} \nonumber\\
		\mbox{win\_free(name)} &\to {\rm\; bool} \nonumber\hspace{-7mm}
	\end{empheq} 
\end{subequations}
Note that topology is not specified in these primitives, and the global topology setting will be taken by  default.

\begin{figure}
    \centering
    \includegraphics[width=0.8\linewidth]{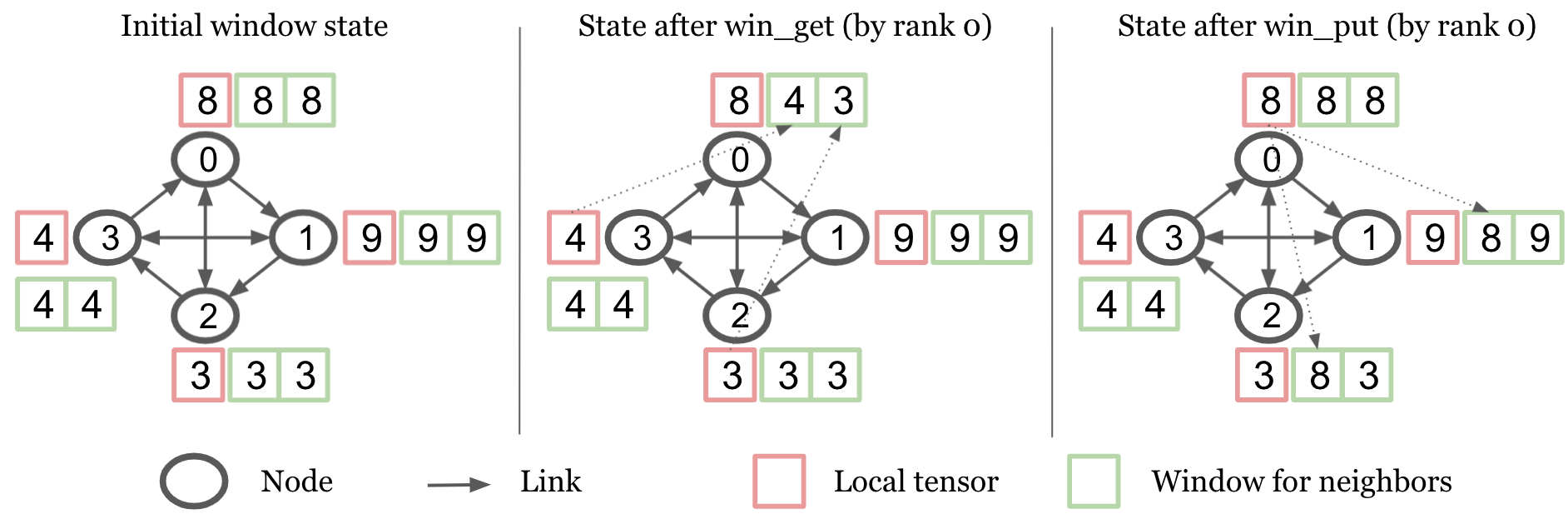}
    \caption{A illustration figure of win\_put and win\_get called by rank 0 (without any dst\_weight or src\_weight setting). In this example figure, the tensor is simply a scalar and each node has two in-coming neighbors.}
    \label{fig:win_ops}
\end{figure}

BlueFog next provides three communication primitives {\ttfamily neighbor\_win\_put, neighbor\_win\_get} and {\ttfamily neighbor\_win\_accumulate} (the names are borrowed from MPI protocol) to manipulate the remote memory. As their name indicate,  {\ttfamily neighbor\_win\_put} puts local tensors to the window buffers maintained by its neighbors while {\ttfamily neighbor\_win\_get}  fetches neighbors' local tensors to its own local window buffers, see the illustration in Fig.~\ref{fig:win_ops}. Primitive {\ttfamily neighbor\_win\_accumulate} performs similarly to {\ttfamily neighbor\_win\_put}, but the former adds local tensor to the existing window buffers maintained by its neighbor while the latter overwrites those buffers. During the window creation, we utilize static topology since frequently allocating and de-allocating windows are expensive. BlueFog provides the dst\_weights and src\_weights dictionary as arguments to the following three primitives to enable asynchronous data transferring defined over the dynamic topology. 
%we adopt the same approach as the synchronous primitive -- providing the dst\_weights and src\_weights dictionary as argument.
Note the window allocation is associated with the global static topology, which implies the ranks used in dst\_weights and src\_weights should be the subset of the neighbors defined under the global static topology.
\begin{subequations}
	\begin{empheq}[box=\widefbox]{align}
	\mbox{neighbor\_win\_get(tensor, name, [src\_weights])} &\to {\rm\; bool} \nonumber\hspace{-7mm}\\
	\hspace{-7mm}\mbox{{neighbor\_win\_put(tensor, name, [self\_weight, dst\_weights])}} &\to {\rm\; bool}  \nonumber\hspace{-7mm} \\
	\hspace{-7mm}\mbox{neighbor\_win\_accumulate(tensor, name, [self\_weight, dst\_weights])} & \to {\rm\; bool} \nonumber\hspace{-7mm}
	\end{empheq} 
\end{subequations}
%\begin{gather}
%    \boxed{
%        \mbox{{neighbor\_win\_put(tensor, name, [self\_weight, dst\_weights])}} \to {\rm\; bool}
%    }\nn\\
%    \boxed{
%        \mbox{neighbor\_win\_get(tensor, name, [src\_weights])} \to {\rm\; bool} 
%    }\nn\\
%    \boxed{
%        \mbox{neighbor\_win\_accumulate(tensor, name, [self\_weight, dst\_weights])} \to {\rm\; bool}
%    }\nn
%\end{gather}
Note that these primitives support either dst\_weights or src\_weights but not both. Also, the put and accumulate primitive are suitable for push-style communication while the get one is for pull-style communication. 

These remote tensor transferring primitives alone are insufficient for asynchronous decentralized communication because 1) the local process does not know when the window buffer is updated and 2) above three communication primitives only transfer or manipulate the tensor within the window buffer, which is not the same tensor that used for local computation. As a result, one more primitive is provided in BlueFog to fill the gap -- {\ttfamily win\_update}. The functionality of {\ttfamily win\_update} is in two-folds. First, it updates the buffer to ensure the tensors kept in window buffers, which may have be changed through {\ttfamily neighbor\_win\_put}, {\ttfamily neighbor\_win\_get}, or {\ttfamily neighbor\_win\_accumulative}, are synchronized and visible to local process. Second, it returns a tensor achieved by averaging the local and neighbors' tensors.
\eq{
    \boxed{
        \mbox{win\_update(name, [self\_weight, src\_weights])} \to {\rm\; tensor}
    }\nn
}
The behavior of {\ttfamily win\_update} operation should return a weighted average tensor based on the local tensor and the latest tensor value from neighbors stored in the local window. We leave the usage of above asynchronous primitives in the asynchronous push-sum example in Sec.~\ref{subsec.async-push-sum}.

\section{Application Examples}\label{sec-application}
BlueFog implemented all previous mentioned communication primitives. Before explanation of implementation and the system design for deep learning usage, this section will first illustrate the usage of the decentralized communication primitives with examples in classical optimization. 

% The first step before implementing decentralized algorithms is to import BlueFog and initialize it.
% \vspace{2mm}
% \begin{lstlisting}[language=python, numbers=left, captionpos=b, label={lst.DGD-static}]
% import bluefog.torch as bf
% bf.init()
% \end{lstlisting}

\subsection{Partial averaging over static graphs for linear regression}

\noindent \textbf{Decentralized linear regression.} Suppose all nodes collaborate to solve the following linear regression problem
\begin{align}\label{dlr}
x^\star = \arg\min_{x\in \RR^d}\quad \frac{1}{2n}\sum_{i=1}^n\|A_i x - b_i\|^2
\end{align}
where $A_i$ and $b_i$ are local data kept by node $i$. The target is to let each node $i$ achieve the optimal solution $x^\star$.

\vspace{1mm}
\noindent \textbf{Decentralized gradient descent.} Decentralized gradient descent (DGD) is among the most widely-used approaches to solving problem \eqref{dlr}. In particular, DGD with static graph topology will iterate as follows: 
\begin{align}
	x_i^{(k+\frac{1}{2})} &= x_i^{(k)} - \gamma A_i^T(A_i x_i^{(k)} - b_i)  \hspace{2cm} \mbox{(local update)} \label{dgd-1}\\
	x_i^{(k+1)} &= w_{ii} x_i^{(k+\frac{1}{2})} + \sum_{j \in \cN(i)} w_{ij} x_j^{(k+\frac{1}{2})}  \hspace{1.3cm} \mbox{(partial averaging)} \label{dgd-2}
\end{align}
where $\cN(i)$ is the in-coming neighbors of node $i$. 

\vspace{1mm}
\noindent \textbf{Code.} We set the topology as the static exponential graph in the following DGD implementation, which is established in \cite{ying2021exponential} to be both sparse and well-connected. Note that such static exponential graph and its associated weight matrix $W$ has already been implemented in the BlueFog library. Users can directly set it 
as the default topology over which DGD runs. The code snippet of the DGD implementation using BlueFog is shown in Listing \ref{lst.DGD-static}. The complete code can be referred to BlueFog online tutorial\footnote{\url{https://github.com/Bluefog-Lib/bluefog-tutorial/tree/master/Section\%203}}. It is observed that the DGD implementation using BlueFog is pretty strait-forward; the code is basically the python interpretation of the math equation \eqref{dgd-1}--\eqref{dgd-2}. In addition to the exponential graph, many other directed or undirected static topologies have also been implemented in BlueFog such as ring, star, mesh, and fully-connected topology. Users can also set up their own topology in BlueFog to facilate decentralized algorithms. 

\vspace{2mm} 
\begin{lstlisting}[language=python, numbers=left, captionpos=b, label={lst.DGD-static},
caption={\small BlueFog implmentation of the DGD algorithm to solve decentralized linear regression.}]
import bluefog.torch as bf
bf.init()  # Initialize the BlueFog

# Set topology as static exponential graph.
G = bf.ExponentialTwoGraph(bf.size())
bf.set_topology(G)

# DGD implementation
for ite in range(maxite):
    grad_local = A.t().mm(A.mm(x) - b)  # compute local grad
    y = x - gamma * grad_local          # local update
    x = bf.neighbor_allreduce(y)        # partial averaging
\end{lstlisting}

Other well-known algorithms such as Exact-Diffusion and Gradient-Tracking to solve problem \eqref{dlr} over static undirected or time-varying directed topology are implemented using BlueFog in Appendix \ref{app-exact-diffusion} and \ref{app-push-sum}. 

\subsection{Partial averaging over time-varying graphs}
\label{subsec:mobile}
\noindent \textbf{Mobile adaptive networks.} Mobile adaptive networks are classical applications of decentralized optimization over networks. 
% In a mobile adaptive network, a collection of nodes with learning and motion abilities will interact with each other locally in order to solve a global distributed processing and distributed inference problems in real-time.
A typical example of mobile adaptive network is the fish schools. While each single fish has limited abilities, the fish schools can exhibit sophisticated behavior that arises from interactions among adjacent members of the school. 
% In fish schools, the individual members tend to move coherently while avoiding collisions. 
When a predator is sighted or sensed, the entire school of fish
adjusts its configuration to disperse. The fish schools can even encircle or attack the predator with remarkable disciplines.

In this subsection, we simulate the above behaviour of fish schools with BlueFog. We mimic each fish with one process in a CPU cluster. With the system-level neighbor-allreduce,  neighbor-allgather, and the topology-oriented APIs provided by BlueFog, we can easily schedule the time-varying fish topology, recognize the dynamic neighborhood, and conduct local information exchange between neighboring fishes. Note that \emph{the topology of fish schools is highly dynamic} especially when fishes are escaping or encircling, this simulation will test how robust BlueFog is to dynamic topology scheduling.

\vspace{1mm}
\noindent \textbf{Problem formulation.} Consider $n$ fishes distributed over some spatial region. Two
fishes are said to be neighbors if they are within a predefined distance. The objective
of the fish school is to estimate the predator's location in a fully decentralized manner
and take actions such as disperse or encircle. The estimation of the predator's location can be formulated into a distributed optimization problem. For fish $i$ at time $k$, it will have a local estimate of the distance $d_i^{(k)}$ and the azimuth angle $\theta_i^{(k)}$ between the predator and itself. If we denote the direction vector as $u_i^{(k)} = [\cos \theta_i^{(k)} \ \sin \theta_i^{(k)}]^T \in \mathbb{R}^2$, the relation between distance $d_i^{(k)}$, the fish's current position $x_i^{(k)}$, and the predator's position $w^\star$ can be characterized as $d_i^{(k)} = (u_i^{(k)})^T (x_i^{(k)} - w^\star) + n_i^{(k)}$,
% \begin{align}
% d_i^{(k)} = (u_i^{(k)})^T (x_i^{(k)} - w^\star) + n_i^{(k)}
% \end{align}
where $n_i^{(k)}$ denotes additive noise. If we let $f_i(w) = \frac{1}{2}[d_i - u_i^T(x_i - w)]^2$ to be the local loss function of fish $i$ to estimate the predator's position $w^\star$, the global loss function of the fish schools to estimate the predator's position can be formulated as $w^\star = \arg\min_w\{\sum_{i=1}^n f_i(w)\}$,
which is a distributed optimization problem and can be solved with decentralized stochastic gradient descent  \eqref{dsgd-1} and \eqref{dsgd-2}. When $w_i^{(k+1)}$ is estimated at time $k$, each fish will either escape from the predator or encircle it at time $k+1$. The detail to formulate these fish behaviors can be referred to \cite{tu2011cooperative}. 

\vspace{1mm}
\noindent \textbf{Code.} The simulation of fish schools with BlueFog is to illustrate how to use partial averaging over time-varying graphs. The code snippet is shown in Listing \ref{lst:fish-school}.
Note the neighborhood at each iteration is determined by the argument {\it src\_weights}, which is a dictionary mapping the rank to the scaling weights.  The argument {\it src\_weights} is updated at each iteration through the neighbor location collections function and Metropolis-Hastings Rule.

\begin{figure*}[t!]
    \centering
    \includegraphics[width=0.23\textwidth]{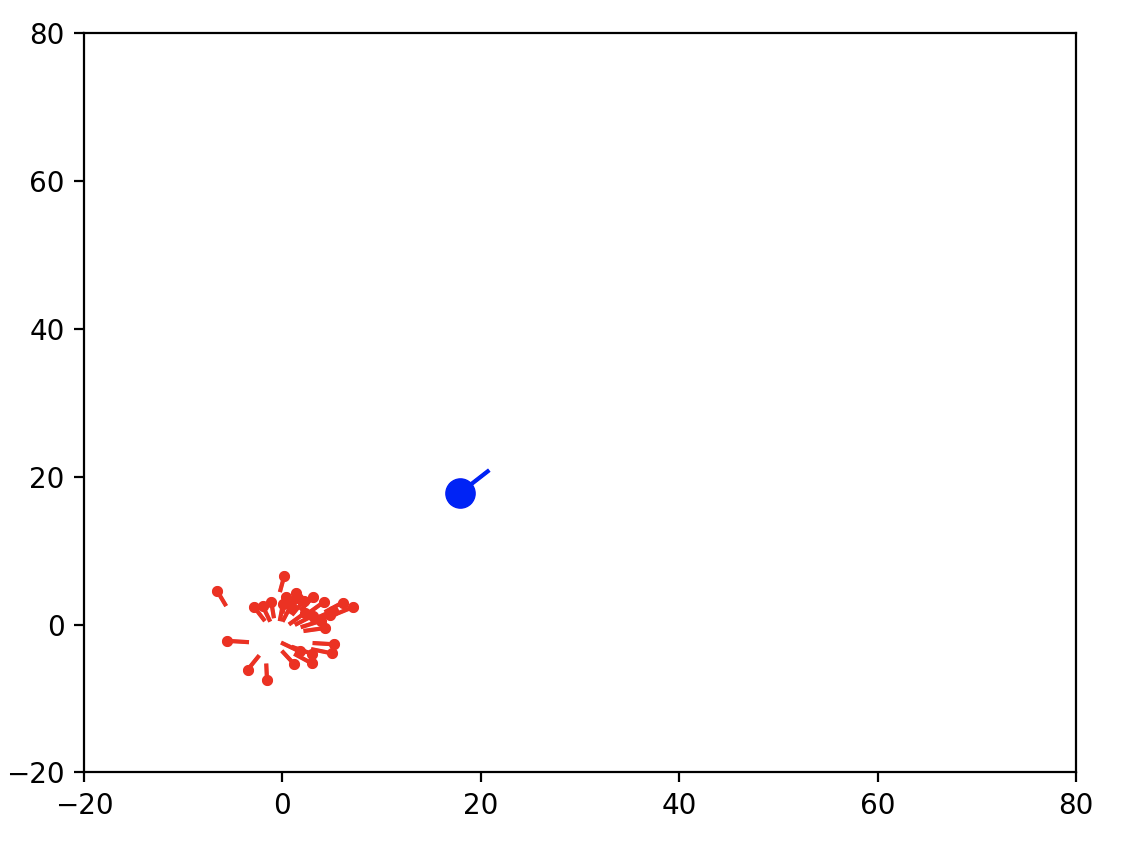}
    \includegraphics[width=0.23\textwidth]{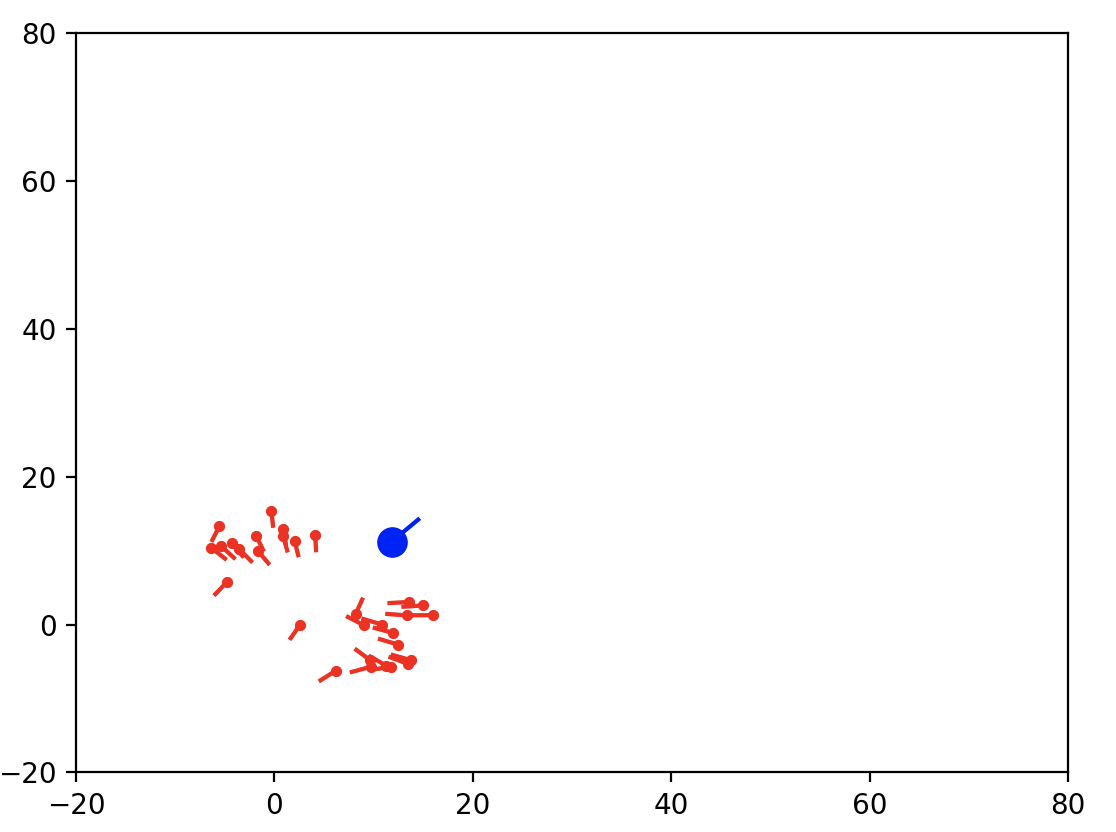}
    \includegraphics[width=0.23\textwidth]{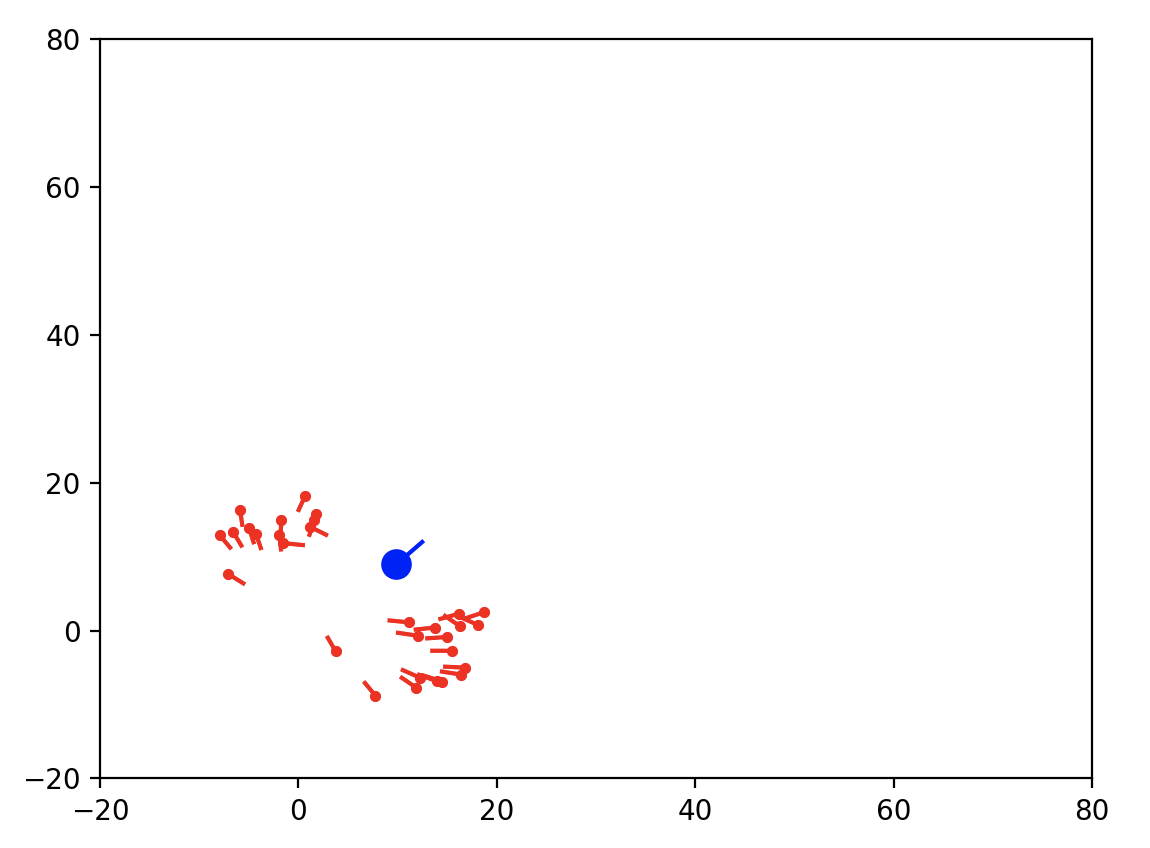}
    \includegraphics[width=0.23\textwidth]{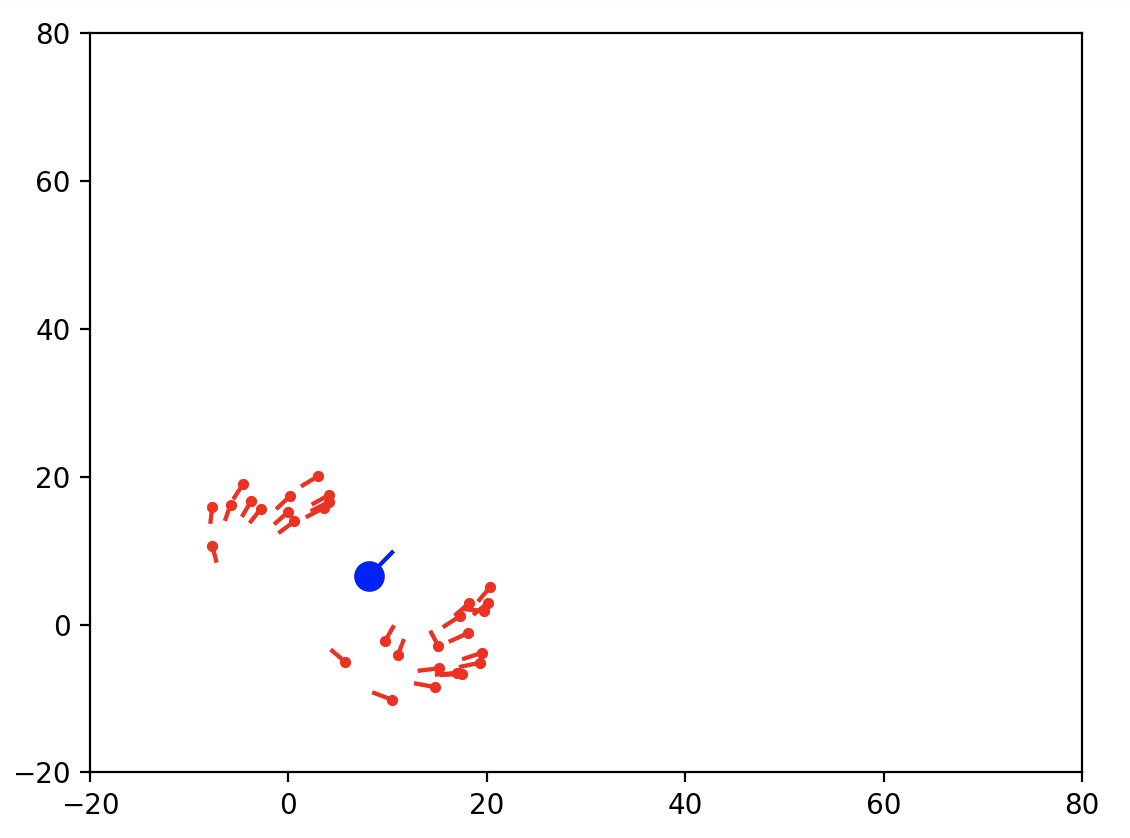}
    \caption{\small The fish schools dispersed when encountering a predator.}
    \label{fig:fish-disperse}
\end{figure*}

\begin{figure*}[t!]
    \centering
    \includegraphics[width=0.23\textwidth]{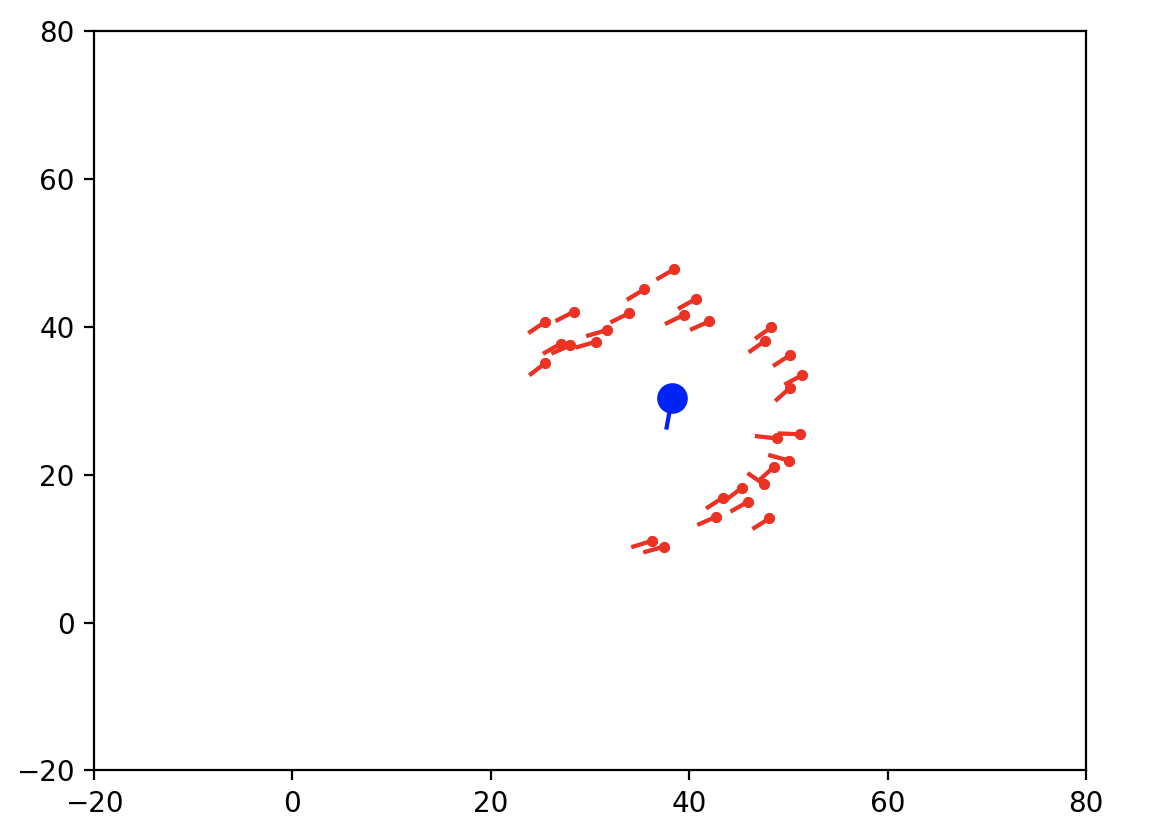}
    \includegraphics[width=0.23\textwidth]{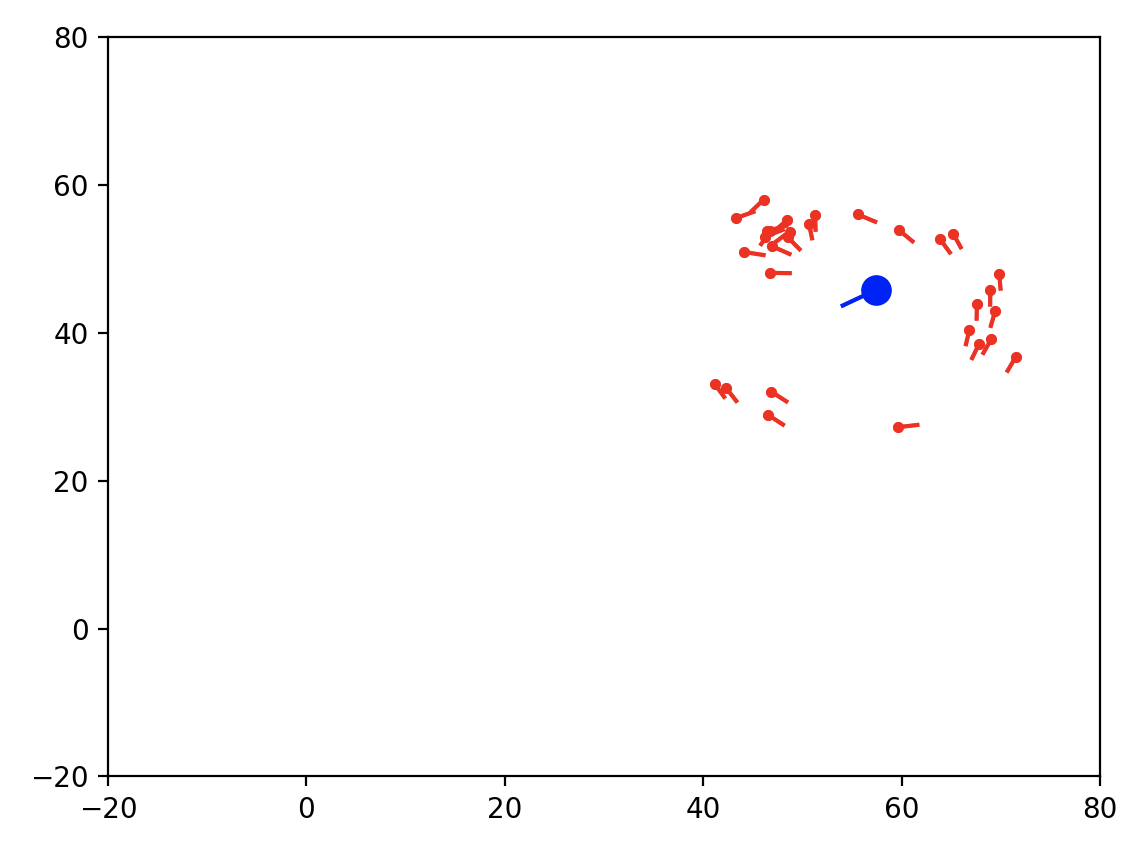}
    \includegraphics[width=0.23\textwidth]{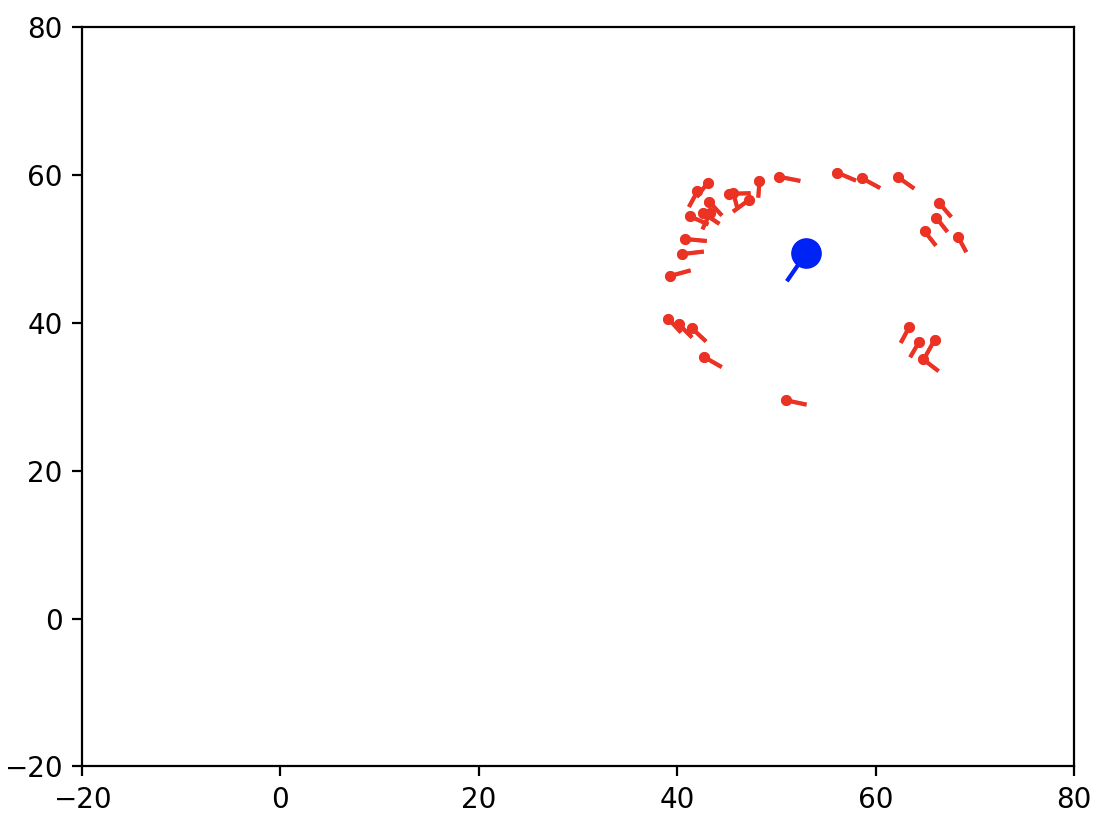}
    \includegraphics[width=0.23\textwidth]{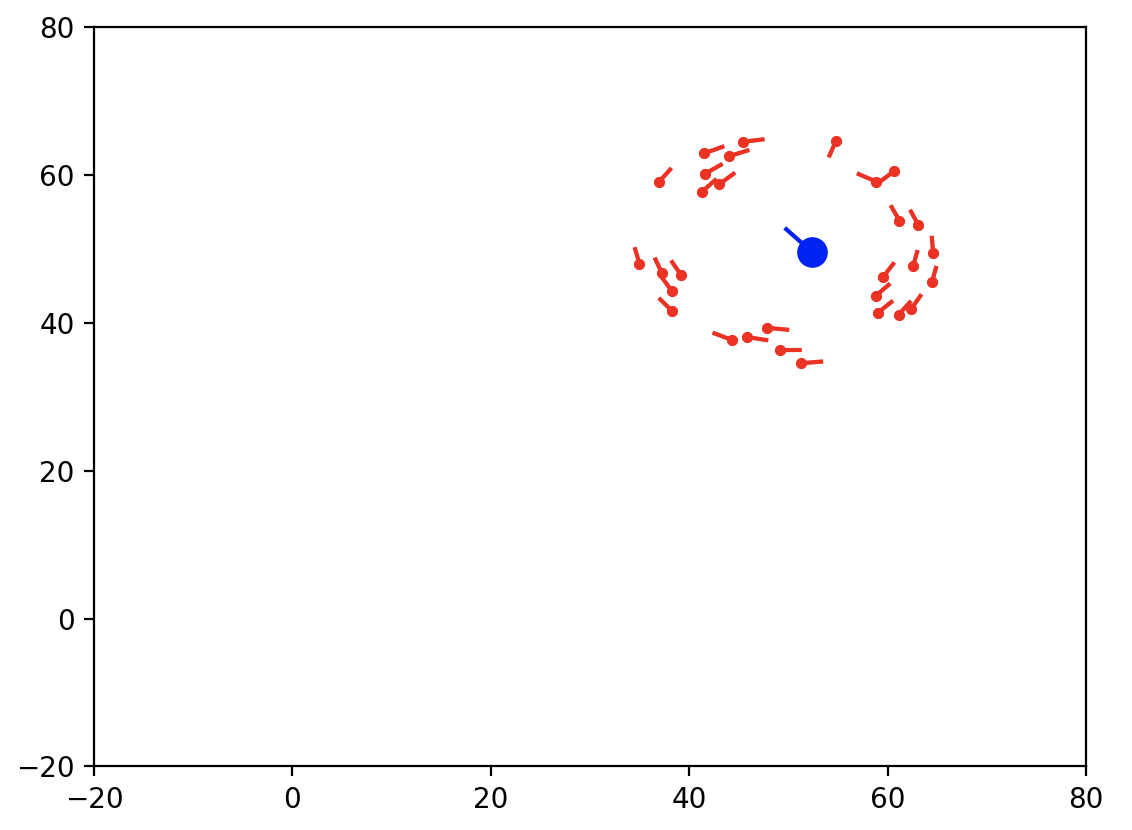}
    \caption{\small The fish schools encircled and trapped the predator.}
    \label{fig:fish-encircle}
\end{figure*}

\begin{lstlisting}[language=Python, caption=Code snippet to simulate school fishes, label={lst:fish-school}]
# x: self location; v: self velocity; w: predator location;

for k in range(maxite):
    ...
    nb_ranks, nb_degrees = get_neighbor_ranks(x, loc_map, threshold)
    
    # Set topology weights according to Metropolis-Hastings Rule.
    self_weight, src_weights = get_mh_weights(nb_ranks, nb_degrees, bf.rank())
    
    # observe noisy dist and direct to predator
    # w_star is a predefined predator's position
    # w_star can change with time
    d, u = get_dist_and_direct(x, w_star)
    
    # estimate predator's position w with D-SGD
    grad = cal_grad(x, w, d, u)
    w = w - gamma * grad
    # Pull-style neighbor communication
    w = bf.neighbor_allreduce(x, self_weight=self_weight, src_weights=src_weights) 
    
    # take escape or encircle actions
    # location and velocity will get changed
    if escap_cond:
        x, v = escape(x, v, w)
    elif encircle_cond:
        x, v = encircle(x, v, w)
\end{lstlisting}
\noindent \textbf{Simulation results.} Figs. \ref{fig:fish-disperse} and \ref{fig:fish-encircle} depict how the fish schools disperse from and encircle a predator, respectively. These two figures illustrate that BlueFog is very effective to conduct partial averaging over dynamic topologies. 
% \begin{figure}[h!]
% 	\centering
% 	\includegraphics[width=0.305\textwidth]{images/fish-disperse1.png} 
% 	\includegraphics[width=0.3\textwidth]{images/fish-disperse2.png}
% 	\caption{\color{red} \small The fish schools dispersed when encountering a predator. The Blue fish denotes the predator.}
% 	\vspace{-2mm}
% 	\label{fig:fish-disperse}
% \end{figure}

% \begin{figure}[h!]
% 	\centering
% 	\includegraphics[width=0.3\textwidth]{images/fish-encircle1.png} 
% 	\includegraphics[width=0.315\textwidth]{images/fish-encircle2.png}
% 	\caption{\small The fish schools encircled and trapped the predator.}
% 	\label{fig:fish-encircle}
% 	\vspace{-2mm}
% \end{figure}

\subsection{Asynchronous push-style partial averaging} \label{subsec.async-push-sum}

\noindent \textbf{Average consensus problem.} This subsection will provide an example on how to use asynchronous push-style primitives. Due to the complexity of asynchronous operations, we only consider the simple average consensus problem for illustration. Suppose every agent $i\in \cV$ has a vector $x^{(0)}_i$. The goal is for all the agents to  obtain the average
\begin{align}
x^\star=\frac{1}{n}\left( x^{(0)}_1+x^{(0)}_2+\dots+x^{(0)}_n \right).
\end{align}
Since the channel between each pair of connected nodes may vary drastically in communication efficiency,  it is possible that some node will finish the partial averaging earlier than the others. Asynchronous average consensus algorithm is proposed to avoid idleness in the fast node when it waits for the slow nodes. 

\vspace{1mm}
\noindent \textbf{Asynchronous push-sum average consensus algorithm.} The vanilla asynchronous average consensus algorithm will result in a biased average for each node. To remove the bias, the push-sum asynchronous algorithm is proposed: 
\eq{
    x^{(k)}_i =& w_{ii} x^{(k-1)}_i + \sum_{j \in \cN(i)}w_{ij} x^{(k-\delta_j)}_j \\
    p^{(k)}_i =& w_{ii} p^{(kt-1)}_i + \sum_{j \in \cN(i)}w_{ij} p^{(k-\delta_j)}_j\\
    y^{(k)}_i =&  x^{(t)}_i /  p^{(k)}_i
}
where $p_i$ is a scalar correct $x_i$ and is initialized as value 1, and $\delta_j$ stands for some index difference since there is no global synchronized index in the asynchronous setting.  When the weight matrix $W = [w_{ij}]$ is column stochastic, i.e. $\sum_{i} w_{ij} = 1$, it is proved in \cite{nedic2014distributed, benezit2010weighted} that each $y^{(k)}_i $ will converge to the unbiased consensus value $x^\star$. For this reason, we will utilize the asynchronous primitives provided by BlueFog in the push-style. 
%One assumption there is the weighted matrix $[w_{ij}]$ is column stochastic, i.e. $\sum_{i} w_{ij} = 1$. Above algorithm can be easily written by BlueFog:

\vspace{1mm}
\noindent \textbf{Code.} The following code snippet is to illustrate how to use partial averaging in the asynchronous and push-style mode. The complete code and experimental results can be referred to BlueFog online tutorial\footnote{\url{https://github.com/Bluefog-Lib/bluefog-tutorial/tree/master/Section\%206}}.

\vspace{2mm}
\begin{lstlisting}[language=python, numbers=left, captionpos=b, label={lst.async_pus-sum},
caption={\small Example of using asynchronous primitives to implement the push-sum algorithm.}]
p = torch.DoubleTensor([1.0]).to(x.device)  # associated p for correction 
x_ext = torch.cat([x, p], 0)
bf.win_create(x_ext, name="x_ext", zero_init=True)

# Set up the weights through the push-style.
outdegree = len(bf.out_neighbor_ranks())
dst_weights = {rank: 1.0 / (outdegree + 1) for rank in bf.out_neighbor_ranks()}
self_weight = 1/(1+outdegree)

for i in range(args.max_iters):
    bf.win_accumulate(x_ext, name="x_ext", self_weight=self_weight,
                      dst_weights=dst_weights, require_mutex=True)
    bf.win_update_then_collect(name="x_ext")
    x, associated_p = x_ext[:-1], x_ext[-1]

bf.barrier()  # Because different processes may end in different time. 
bf.win_update_then_collect(name="x_ext")
bf.win_free(name="x_ext")
\end{lstlisting}

\noindent \textbf{Remark.} There are a few details in Listing \ref{lst.async_pus-sum} that we do not discuss in the previous sections. First, we add a new argument {\ttfamily require\_mutex=True} in {\ttfamily bf.win\_accumulate}. Such mutex is to guarantee the memeory read and  manipulation operations will not be conducted at the same time, which is crucial to avoid the date race problem since two process do not synchronize with each other. Second, we use {\ttfamily win\_update\_then\_collect} instead of {\ttfamily win\_update}. The former operation will reset the window memory, i.e. setting all the elements in the window to be zero, after reading it. This operation needs to be atomic so that the sum of $p_{i}^{(k_i)}$ (the sum of the value in the local tensor and the corresponding windows) over all agents is guaranteed to remain as the initial value in all iterations.

\section{System Design} \label{sec:system-design}
Besides the low-level communication primitives discussed previously, more system-level design is required to construct a powerful framework supporting generic decentralized training with great performance. In this section, we will discuss what extra system-level components and optimizations we deploy enables BlueFog to achieve the state-of-the-art performance for large-scale deep neural network training tasks. 

The design philosophy for these high-level APIs shift a little from these low-level communication primitives. The later ones expose many but universal control knobs out so that the user can have full freedom and capability to design various decentralized algorithms. In contrast, we make high-level APIs concise and easy-to-use to support a few off-the-shelf high performance decentralized optimization solutions.
%Under the hood, it already applied several optimizations for the performance so that the user do not need to worry about them.
But we still leave users the whole control over topology choices which can applied on any generic decentralized algorithm.

First, we provide an example of using BlueFog for a deep neural network training task:
% \begin{minipage}{0.8\linewidth}
\begin{lstlisting}[language=python, numbers=left, captionpos=b, label={lst.dist_opt},
caption={\small Example of using distributed optimizer wrapper. The topology information is passed down to optimizer through external control. In the example, it also periodically runs global allreduce every 20 iterations.}]
import torch
import bluefog.torch as bf
# initialization and model and data settings.
...
opt = optim.SGD(model.parameters(), lr=0.01)
# BlueFog wraps the standard optimizer to use the decentralized communication
# to fuse the information with neighbors.
opt = bf.DistributedAdaptThenCombineOptimizer(
        optimizer=opt, 
        model=model, 
        communication_type=CommunicationType.neighbor_allreduce)
for batch_idx, (data, target) in enumerate(train_loader):
    ... 
    # Generate dynamic topology weights
    opt.self_weights = self_weights
    opt.src_weights = src_weights
    opt.enable_topo_check = True
    # Periodic global allreduce
    opt.communication_type = CommunicationType.allreduce if batch_idx % 20 == 0
                             else CommunicationType.neighbor_allreduce
    # Forward and backward computation.        
    output = model(data)
    loss = F.cross_entropy(output, target)
    loss.backward()  
    opt.step()
    ...
\end{lstlisting}

In the code snippet above, we omit most initialization, model, and data setting. This is because the BlueFog optimizer should be non-intrusive by design and all parts unrelated to the optimizer should remain the same as the standard distributed training code. Basically, what the user need to do is just to wrap the standard optimizer with the BlueFog API, transforming the original optimizer into a decentralized one. As seen in the above example, the wrapped optimizer also accepts the arguments like {\ttfamily src\_weights}, {\ttfamily communication\_types}, etc, to configure the communication to be executed at each iteration.
%, the optimizer may execute different types of communication.
These settings should be placed before the forward and backward computation because the communication may trigger during the forward propagation, the backward propagation, or the step function, depended on what kind of decentralized optimizer is used. Next, we delve into a few key components that BlueFog adopts under these high-level APIs.
% In this example, we use the so-called {\ttfamily DistributedAdaptThenCombineOptimizer}, BlueFog  provides more options
% \end{minipage}

% \subsection{System and API Design }
% With above discussion, the designing of API will follow on the these th
% \begin{itemize}
%     \item {\bf Non-intrusive (high-level)}: The API must be non-intrusive to
% applications. Application developers usually start from
% writing local training scripts, and scale out when hitting the resource limit on a single machine. At that
% point, it is unacceptable to ask developers to rewrite
% the entire application to enable distributed data parallel training. Instead, the developer should be able to
% reuse the local training script with minimal modifications.
%     \item {\bf Interceptive (high-level)}: The API needs to allow the implementation to intercept various signals and trigger appropriate algorithms promptly. Distributed data parallel
% aims at accelerating training by using more computational resources. This process requires subtle optimizations in both computations and communications
% to achieve the best performance. Hence, the API must
% expose as many optimization opportunities as possible
% to the internal implementation.
%     \item {\bf local information only (low-level)}
%     \item {\bf flexible and dynamic (low-level)}
% \end{itemize}
% \paragraph{Topology related APIs}
% \paragraph{Low-level communication APIs}
% \paragraph{High-level optimizer wrapper APIs}

% \subsection{System-level optimizations}

\subsection{Overlapping communication and computation} \label{subsect.overlap}
Typically, both communication and computation are time-consuming resource-intensive operations. The communication operations are mainly handled by network interface cards (NICs) and the communication overhead mostly comes from waiting the information sending to / received from other processes, while most (forward or backward propagation) computation operations happen on the CPU/GPU. These two types of operations by nature are independent, so ideally, executing communication and computation in parallel whenever possible is a good strategy to reduce overhead and make full use of system resources, which is also the key to gain more linear scaling speedup in the decentralized learning scenario. 

Here, we first focus on the low-level primitives, which allows users not only to design various decentralized algorithms freely but also to easily write programs enjoying great performance boost by parallelism. In the later sub-section, we will show how we integrated this notion into the decentralized optimization algorithm to provide system level optimization. 

To empower this idea, we support the non-blocking version of {\ttfamily neighbor\_allreduce} primitives. The nonblocking version shares the same input arguments with the blocking counterpart, but they differ in what they return. The blocking one wait until the communication is finished then returned the combined tensor, which obviously doesn't allow the user to overlap it with computation. The nonblocking one, instead, returns a handle (basically an unique integer) to identify the communication request. Because the nonblocking API doesn't need to wait for other nodes to finish communication, it can return the handle immediately. Behind the scene, BlueFog system transfers the handle to a separate thread dedicated for communication. Hence, the user can write other computation routine after the nonblocking API. When the results of communication are required, an extra  {\ttfamily bf.wait()} API, taking the previous returned handle as input, should be called. It mainly synchronize with other nodes to ensure the communication finishes, returning a reduced tensor as output. The following Listing \ref{lst.nonblocking} illustrates the usage of non-blocking operation.

\begin{lstlisting}[language=python, numbers=left, captionpos=b, label={lst.nonblocking},
caption={\small Example of using nonblocking function to overlapping the communciation and computation.}]
handle = bf.neighbor_allreduce_nonblocking(x, self_weight, src_weights, dst_weights)
grad = ComputeGradient(x)  # Communication in parallel
# Wait() returns the tensor until the communication is done
x = bf.wait(handle) - lr * grad
\end{lstlisting}

As for the asynchronous operations, like {\ttfamily neighbor\_win\_put, neighbor\_win\_get}, etc, they also have their corresponding nonblocking version of APIs. Note that asynchronous and nonblocking are two orthogonal concepts here. The former is defined between the two separated nodes or processes and the later is defined within the local node between communication and computation threads.

\subsection{Enabling hierarchical communication}
\begin{figure}[!htbp]
\begin{minipage}[t]{0.35\textwidth}
% \begin{wrapfigure}{L}{0.4\textwidth}
    \centering
    \includegraphics[width=.9\textwidth]{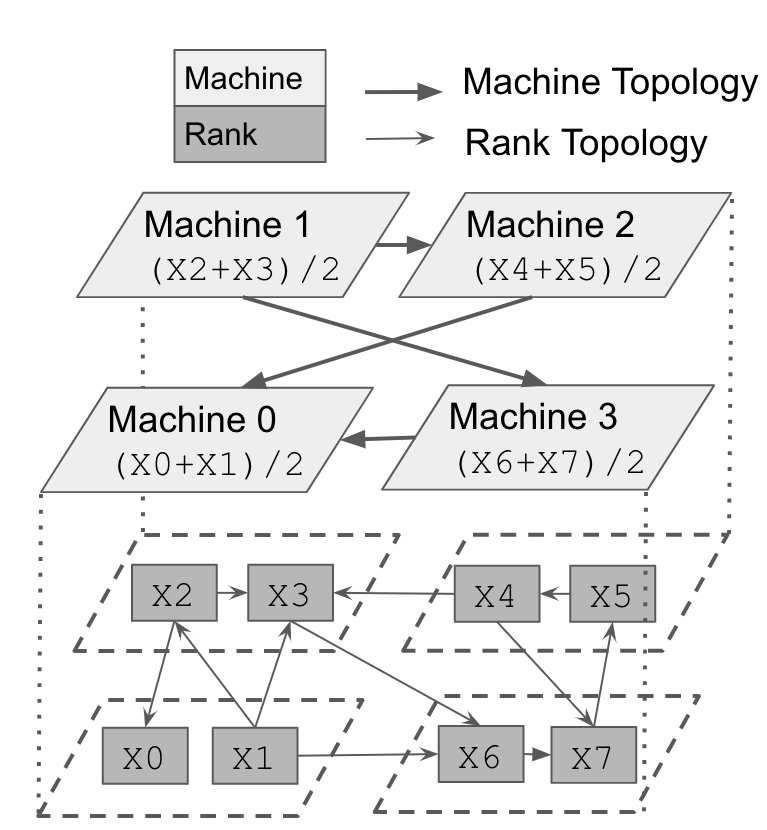}\vspace{-3mm}
    \caption{\small The illustration of {\it hierarchical\_neighbor\_allreduce}. In this example, each machine own two processes and $Xk$ represents the input tensor at rank $k$.  
    %In hierarchical view, each machine is super node associated with machine topology and each machine own the average value of local ranks.
    } 
    \label{fig:h_nar_logic} \vspace{-3mm}
% \end{figure}
\end{minipage}\hspace{3mm}
\begin{minipage}[t]{0.58\textwidth}
% \end{wrapfigure}
% \begin{wrapfigure}{l}{0.4\textwidth}
% \begin{figure}[!tp]
    \centering
    \includegraphics[width=.9\textwidth]{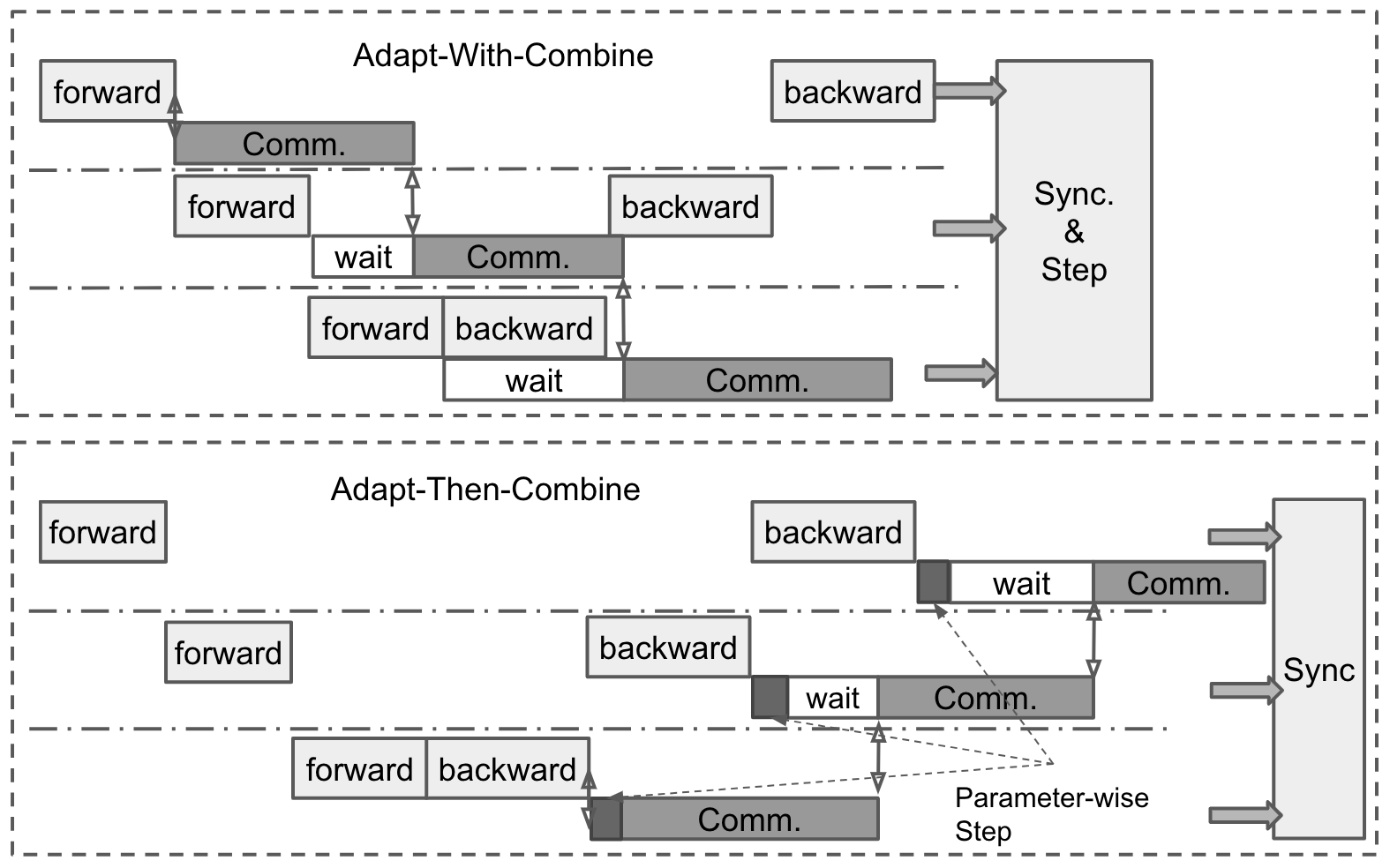}\vspace{-2mm}
    \caption{\small A timeline illustration of AWC and ATC algorithm over a three-layer neural network. Note this is an illustration for the toy example without other optimization techniques. In practice, due the tensor fusions and negation between the nodes, the pattern may vary. }
    \label{fig:timeline}
\end{minipage} 
\end{figure}

Previously, we only consider the ideal case where each node is equivalent to each other; the nodes are isotropic in terms of placement; the bandwidth between each node pair is the same. 
However, %it is rarely true in the practical setting.
it is extremely difficult to have a practical system perfectly align to these ideal conditions. Here we mainly consider one case that the bandwidth and placement of GPUs are different. Take DGX-1 for instance\cite{dgx-1}. Its intra-machine communication between GPUs uses high-speed NV-Links, while its inter-machine communication uses slower NICs. Multiple rounds of intra-machine communication may take the same or even less time than a single inter-machine one. This implies that we should minimize the inter-machine communication as long as it does not change the functionality,

Hence, BlueFog introduces another API {\ttfamily hierarchical\_neighbor\_allreduce} for networks with two tiers of communication speeds, which contain multiple super nodes (machines) benefiting from faster intra-machine communication. See Fig.~\ref{fig:h_nar_logic} as an example. Its execution can be divided into three stages. Firstly, within the same machine, all local nodes communicate with each other and formulate a local averaged tensor to represent the result from the machine. Then, the super node (a particular local node inside the machine) talks with the machine neighbor to compute the neighborhood averaging. Lastly, all the local nodes within the machine set the neighbor averaged value from machine level communication as its local averaging result.

Unlike the hierarchical version of {\ttfamily allreduce} operator, which shares the same functionality but just different implementations with non-hierarchical one, the {\ttfamily hierarchical\_neighbor\_allreduce} is no longer equivalent to its non-hierarchical counterpart in terms of functionality. Mainly, it is because the topology and the corresponding neighborhood definition are based on the super-node (machine) level instead of local node (process rank) level.
% Similar to \textit{neighbor\_allreduce}, we need to configure a global level topology for static usage, and provide the two weight arguments for dynamic usage for \textit{hierarchical\_neighbor\_allreduce}. Note that here the machine rank of the super node is employed instead of the process rank used in \textit{neighbor\_allreduce}. Fortunately,
Under a homogeneous environment, i.e. the number of nodes inside each machine is the same, the translation between node rank and machine rank is easy as follows:\vspace{-1.5mm}
\eq{
    {\rm machine\_rank} =& {\rm rank \;//\; local\_size},\;\;\;\;\; {\rm rank} ={\rm machine\_rank * local\_size + local\_rank} \nn\\[-8mm]\nn
}
where $//$ means the integer division and local\_rank and local\_size represent the rank and size of processes inside one physical machine/super node, respectively.
The behavior of the hierarchical API is ill-defined when difference machines have different number of processes. So, we don't recommend the usage of this API under this situation.
In summary, we have these two new corresponding APIs
% \begin{subequations}
% 	\begin{empheq}[box=\widefbox]{align}
% 	    \mbox{set\_machine\_topology(graph\_object)} \to {\rm\; bool}\nn\\
% 	    \hspace{-8mm}\mbox{hierarchical\_neighbor\_allreduce(tensor, name[, self\_weight, src\_machine\_weights, dst\_machine\_weights])} \!\to\! {\rm\; bool}\hspace{-10mm}\nn
% 	\end{empheq}
% \end{subequations}
\begin{gather}
    \boxed{
        \mbox{set\_machine\_topology(graph\_object)} \to {\rm\; bool}
    }\nn\\
    \boxed{
        \mbox{hierarchical\_neighbor\_allreduce(tensor, name, [self\_weight, src\_machine\_weights, dst\_machine\_weights])} \!\to\! {\rm\; bool}
    }\nn
\end{gather}
Note we adopted the same style as {\ttfamily neighbor\_allreduce} except adding keyword `machine' to the arguments. Be default, the machine rank that each process or node belongs to is automatically detected by the program.
%For ease usage, it is also allowed to set virtual machine rank on each process. It is not useful in the production environment since it may degrade the performance.

%After we have the machine-level values and topology, the rest logic is the same as the neighbor allreduce.

\subsection{Triggering communication in deep learning optimizers}\label{sec:trigger-comm}
With the nonblocking APIs discussed in Section \ref{subsect.overlap}, we are able to overlap the time for communication and computation. Obviously, more time we can overlap between them, better linear speedup we can get when the network size scales up. In consequence, we should trigger the communication as early as possible.
%Just as the decentralized communication,
Typically, the decentralized algorithms have at least two variants in terms of the execution order between communication and computation.  In contrast, the allreduce based algorithm is always triggered after gradients are computed by the backward propagation. Using the Decentralized SGD algorithm as example, we have the following two variants:
\eq{
    x_i^{(k)} =& \overbrace{\sum_{j\in \cN^{\dagger}(i)} w_{ij} x_j^{(k-1)}}^{\rm Communication} - \gamma \overbrace{\nabla F(x_i^{(k)};\bxi_i^{k+1})}^{\rm Computation} &{\rm (AWC\ style)} \label{eq.awc}\\
    x_i^{(k)} =& \underbrace{\sum_{j\in \cN^{\dagger}(i)} w_{ij} \left(x_j^{(k-1)} - \gamma \underbrace{\nabla F(x_i^{(k)};\bxi_i^{k+1})}_{\rm Computation} \right)}_{\rm Communication} &{\rm (ATC\ style)} \label{eq.atc}
}
We call \eqref{eq.awc} as Adapt-While-Communicate (AWC) style algorithm and call \eqref{eq.atc} as Adapt-Then-Communicate (ATC) style algorithm. As their names indicated, AWC algorithm can perfectly perform the computation and communication in parallel while ATC algorithm have to communicate after the computation is done. The situation is slightly more complicated if we apply these two algorithm variants on the deep neural network (DNN) training since gradient computation for DNN is layer-wise \cite{werbos1990backpropagation, rumelhart1985learning}. This layer-wise computation nature implies that we can execute the communication of partial parameters (commonly the parameters in one layer) when all pre-requisition computation is done. This is crucial because even for the ATC-style algorithm (\ref{eq.atc}) 
although the communication has to wait until the gradient is computed, we can still overlap the communication of one layer with the computation of the next layer.  Consider a concrete example of applying \eqref{eq.awc} and \eqref{eq.atc} on a toy three-layer neural network, of which the timeline is illustrated in Fig.~\ref{fig:timeline}. In deep learning frameworks like PyTorch, it provides the hook that can register the function to be triggered when the forward computation or backward propagation of a layer is finished. In the figure, we can see that we register the communication for the AWC-style algorithm when the forward computation is done since it can maximize the overlapping; while it is not possible for the ATC-style algorithm but we can still register it under the backward one. Moreover, it is not hard to deduce that the deeper the neural network is, the larger portion the communication in ATC-style algorithm may overlap with its computation.
Here we discussed the optimizers corresponding to ATC- and AWC-style. Actually, BlueFog provides more off-the-shelf high-level APIs for different styles of optimizers. Please refer to the BlueFog documentation website for more information.

\subsection{Miscellaneous system components}
There are more  system components that BlueFog implements for extra functionalities or performance boost such as tensor fusion for the merging multiple small tensor together, difference scaling strategies for the push-style and pull-style settings, timeline function to analysis the usage of each operation,  the distributed mutex implementation for the window primitives, tolerating different orders of gradients or parameters triggered in different agents, etc. Since BlueFog already provides several wrapped decentralized optimizers off-the-shelf, most users do not need to know these low-level details. If the user want to implement his/her own decentralized algorithm, mainly they just need to focus on what communication primitives to use and when and how to overlap the communication and computations. After these design decisions are made, the user can refer to our preset optimizers and create his/her own optimizer.

\section{Implementation} \label{sec:implementation}
\begin{figure}[!tp]
    \centering
    \includegraphics[width=.6\textwidth]{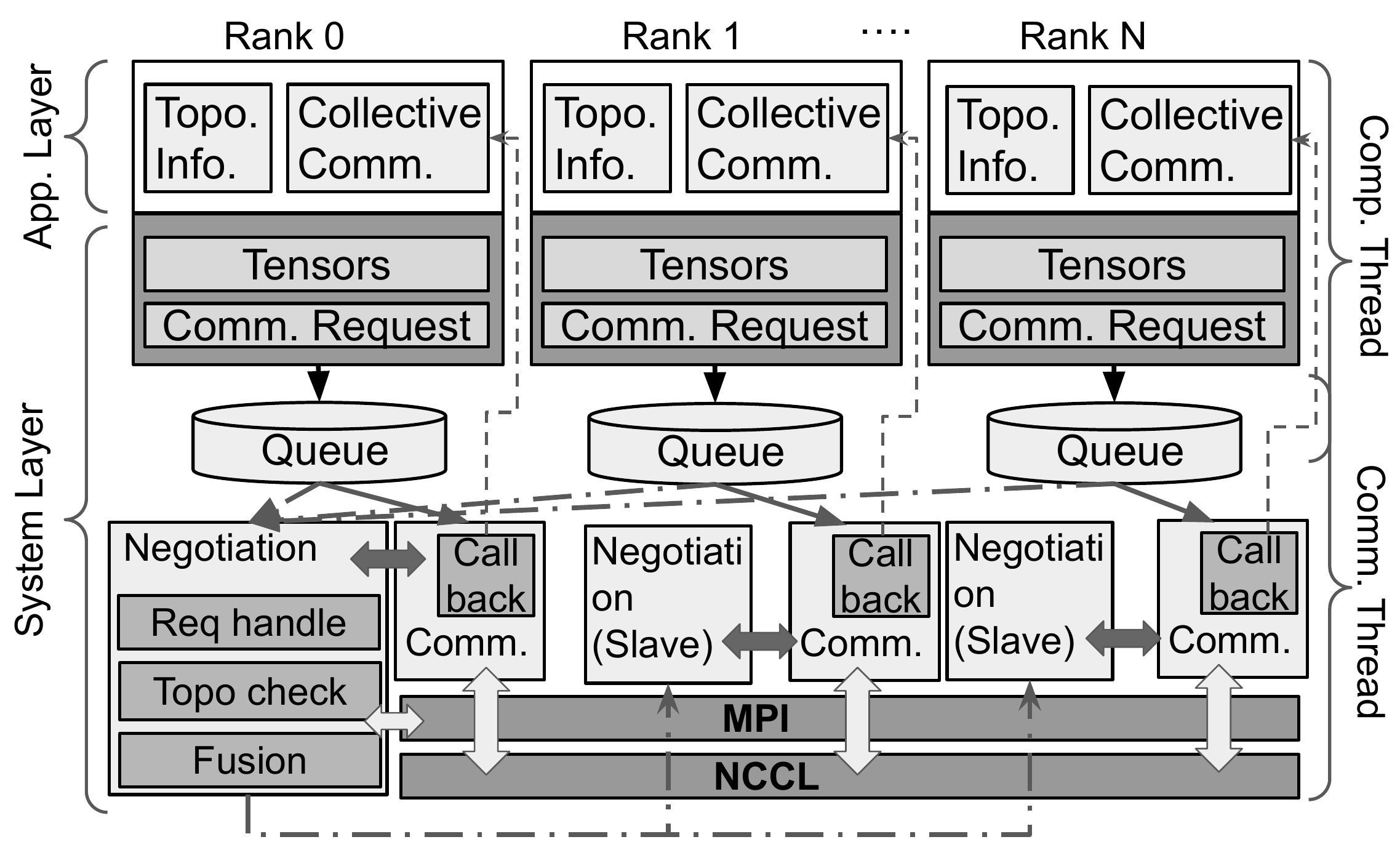}\vspace{-2mm}
    \caption{\small BlueFog's architecture.}
    \label{fig:architecture}\vspace{-3mm}
\end{figure}

BlueFog fully integrates with the PyTorch library \cite{paszke2019pytorch} so that the BlueFog APIs can be easily called in Python for decentralized tensor/vector computation and the core communication logic is implemented in C++ for better efficiency and faster interaction with low-level communication libraries like OpenMPI and NCCL.  It is easy to install BlueFog by simply running {\ttfamily pip install bluefog}.

\subsection{Architecture Overview}
BlueFog architecture is composed of an application layer and a system layer. The former implements the low-level APIs and the optimizer wrapper described in Section \ref{sec:decen_comm_abs} and Section \ref{sec:system-design}. The latter is dedicated to the underneath actual communication protocol in the system. Fig. \ref{fig:architecture} provides an overview of the entire architecture.

The application layer consists of three types of APIs: 1) topology management APIs such as {\ttfamily set\_topology} and {\ttfamily set\_machine\_topology}; 2) low-level communication APIs such as {\ttfamily allreduce}, {\ttfamily neighbor\_allreduce}, {\ttfamily neighbor\_win\_get}, etc.; 3) high-level distributed optimizer wrappers. In addition, BlueFog provides {\ttfamily bfrun}, a thin wrapper over {\ttfamily mpirun} to initiate BlueFog processes, and {\ttfamily ibfrun} to use BlueFog in interactive python environment such as Jupyter Notebook.

The system layer consists of two threads. The main thread, created by Python, is responsible for computation and generating communication requests. After the communication API like {\ttfamily neighbor\_allreduce} is called by application layer, BlueFog transforms them into a proper request structure and push that into the shared queue. The application layer will then receive the handle of the enqueued request for fetching results in the future. Another thread is dedicated to communication, necessary for the nonblocking routine implementation. It pops out the requests from the shared queue, and then negotiates the requests with other BlueFog processes to schedule the communication order, which will be discussed more in Section \ref{sec.impl.nego}. When all processes are ready, the negotiation service will inform the execution service to exchange the tensor information between processes.

\subsection{Implementation of Low-level Primitives}
BlueFog implements the {\ttfamily neighbor\_allreduce} and {\ttfamily hierarchical\_neighbor\_allreduce} on both MPI and NCCL.
Unless specified through the environment variables, BlueFog automatically chooses the underlying communication library. If the tensor is on CPU, MPI takes the responsibility of communication. If the tensor is on GPU, NCCL is used. 
\begin{wrapfigure}{R}{0.48\textwidth}\vspace{-5mm}
    \centering
    \includegraphics[width=.47\textwidth]{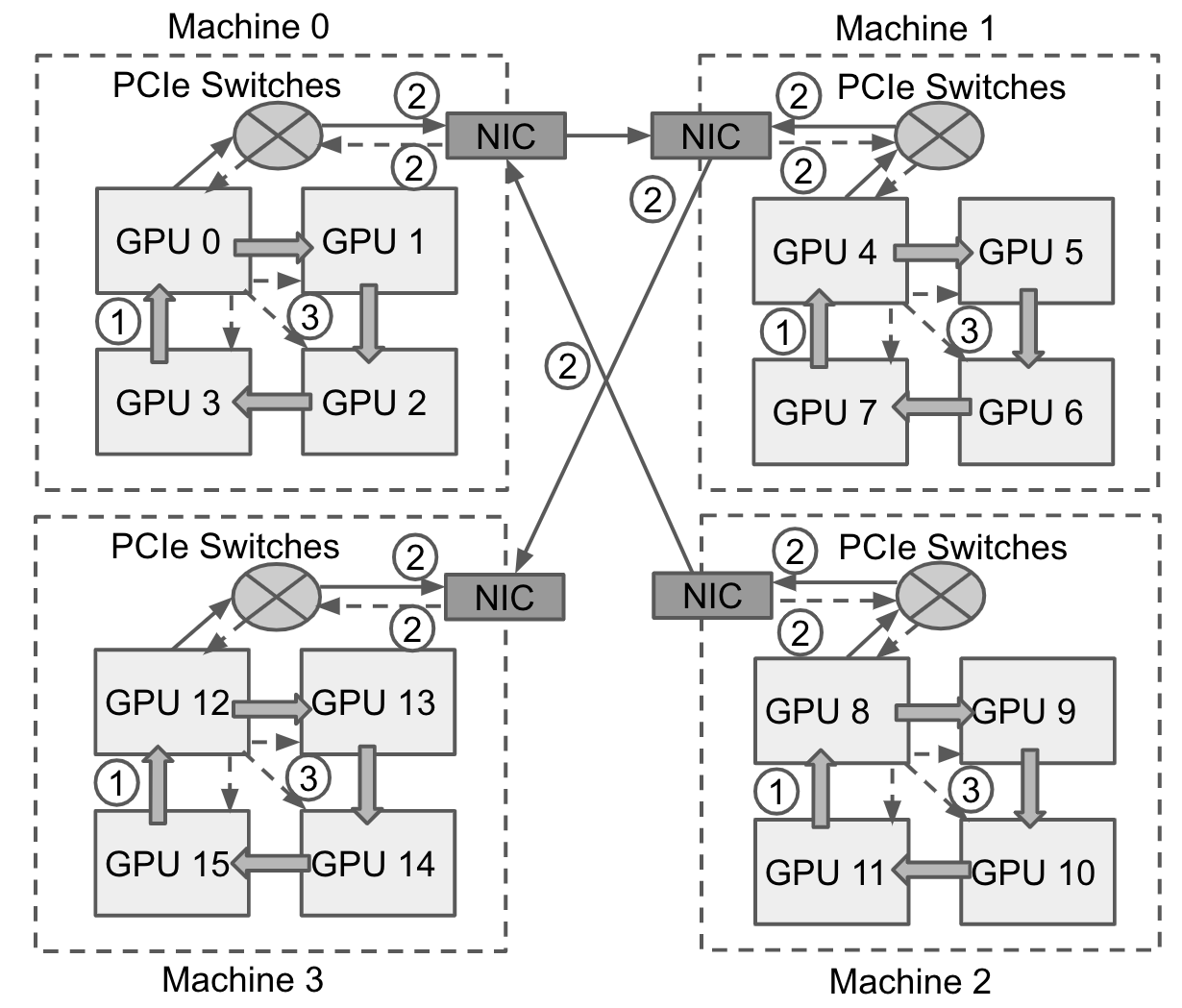}
    \caption{\small An illustration Dataflow of Hierarchical Neighbor Allreduce of 4 GPUs in one machine. 1.  Intra-machine allreduce over NV-Links. 2. Inter-machine neighborhood communication through NIC. Here we skipped the CPU for simplicity. It may need CPU involvement if it running with TCP or without running with RDMA 3. Intra-machine Broadcast of received neighbor information.}\vspace{-1mm}
    \label{fig:h_nar}
\end{wrapfigure}

\paragraph{MPI as backend}
Although only a single synchronous neighborhood communication API is used for both static and dynamic topology usage, the implementation for them is different underneath.

For the static topology, BlueFog relies on neighborhood collective operations, which are added in MPI-3 standard \cite{gropp1999using, gropp2014using, mpi-3.0}.  Among them, the {\ttfamily MPI\_Neighbor\_allgatherv}\footnote{Notice it is allgatherv instead of allgather. The 'v' here representing the varying size since each node may have different number of neighbors so that the gathering information will end up with varying size.} API fits our requirements.
%most closely. }
Roughly, we can build the graph communicator for the provided global static topology. Then, neighbor information is collected through the {\ttfamily MPI\_Neighbor\_allgatherv} routine. Lastly, the weighted average over received information will generate the final {\ttfamily neighbor\_allreduce} result, where the weights can be specified by the {\ttfamily self\_weight}, {\ttfamily src\_weights}, and {\ttfamily dst\_weights}  arguments. The advantage of using the native MPI routine is that MPI vendor may optimize the transmission order in the graph to avoid the traffic conflict as much as possible. 

But this approach is not ideal for the dynamic topology. Creating graph communicators for all possible dynamic topology is expensive. Managing and associating the topology with the graph communicator is difficult because matching two dynamic graphs can be inefficient. Hence, the peer-to-peer communication is used to replace {\ttfamily MPI\_Neighbor\_allgatherv}. The destinations and sources of peer-to-peer communication are established on-the-fly through the  {\ttfamily src\_weights} and {\ttfamily dst\_weights} arguments provided by users. To alleviate the communication congestion that multiple processes send to the same destination, the destination order at each process is sorted based on the difference between its own rank and the destination rank.

To implement {\ttfamily hierarchical\_neighbor\_allreduce}, it is slightly different from the previous implementation for {\ttfamily neighbor\_allreduce}. It roughly takes four steps: 1) Intra-machine allreduce (SUM); 2)  Inter-machine neighborhood communication; 3) Intra-machine broadcast; 4) Reduce the received tensor locally.
% \begin{enumerate}
%     \item Intra-machine allreduce (SUM);
%     \item Inter-machine neighborhood communication;
%     \item Intra-machine broadcast;
%     \item Reduce the received tensor locally.
% \end{enumerate}
The inter-machine communication is always executed by local rank 0 in each machine. To achieve inter- and intra-machine communication, two more communicators are built. One is a local communicator that connects all the processes in one machine and another is a global communicator that connects all processes with local rank 0. For static and dynamic machine topology usage, the inter-machine communication is the similar as {\ttfamily neighbor\_allreduce} mentioned before. 

The asynchronous primitives' implementation is based on the window operations provided after MPI-3 standard. Internally, we maintain a state dictionary that maps from the unique window name to the registered window object. Each window object is also associated with a distributed mutex that can be locked and unlocked by the local process and the corresponding neighbor process. Since each window object may associate with multiple neighbors, each data manipulation primitive will automatically calculate the displacement value for the remote window based on the tensor shape and the rank.

\paragraph{NCCL as backend} NCCL implementation depends on the NCCL version. Before the NCCL 2.7 version \cite{nccl27}, it only \liu{provides} the collective communications -- allreduce/broadcast/allgather, etc. In order to implement {\ttfamily neighbor\_allreduce}, we have to create \liu{multiple} pair communicators and then use broadcast under each pair communicator to mimic the send and receive operations. To avoid potential dead-lock problems, we sort the order of pair communicators first to align the send/receive usage 
among different processes. Each pair communicator is uniquely identified by the source and destination ranks, which are used to determine the order. Fortunately,  NCCL \liu{provides} the peer-to-peer communications since version 2.7. New {\ttfamily ncclSend} and {\ttfamily ncclRecv} APIs are introduced \cite{nccl27}, which enables more efficient implementation and boosts our performance as well. 
We use the group functions {\ttfamily ncclGroupStart} and {\ttfamily ncclGroupStop} merging multiple peer-to-peer communications to form the neighborhood communication. Using these group functions, the deadlock problem is completely handled by NCCL.

As for {\ttfamily hierarchical\_neighbor\_allreduce}, NCCL's implementation is similar as the MPI case, which also takes the same four steps. The advantage of using {\ttfamily hierarchical\_neighbor\_allreduce} over GPU is because it can fully exploit the high-speed NVLink that has been optimized for allreduce operation.
An illustration dataflow of {\ttfamily hierarchical\_neighbor\_allreduce} of 4 GPUs in one machine is shown in Fig.~\ref{fig:h_nar}.  As for the asynchronous primitives, BlueFog haven't fully supported it with NCCL yet. We are interested to fulfill it in the future work.

\subsection{Negotiation Service} \label{sec.impl.nego}
In order to maximize the performance and fault tolerance, BlueFog adopts similar techniques from Horovod \cite{sergeev2018horovod}. Most of these functionalities happen under the negotiation service. During the negotiation step, rank 0 process collects the collective communication requests from all processes. Each request has a unique \liu{associated name} so that the negotiation service can determine if the tensors from all processes are ready or not. This readiness functionality is necessary because the order of execution of tensors may be distinct between different ranks.

Besides the readiness responsibility, it also performs several sanity checks including \liu{whether} the operations are matched or not, \liu{whether} the dynamic topology is valid or not, etc. Topology check is one of the major components among these sanity checks. As discussed in Section \ref{sec:decen_comm_abs}, the primitives used in Bluefog are encouraged to perform in a local view. Therefore, there is chance that users may provide unmatched destination and source information to the primitives. For instance, user fills in {\ttfamily dst\_weights} in process $i$ to push information to process $j$, but does not provide {\ttfamily src\_weights} in process $j$. As a result, process $j$ is agnostic to its in-coming neighbor $i$, which hangs the program. To avoid that, before the heavy tensor communication, the negotiation service synchronizes the ranks of sending and receiving among the entire network to ensure the topology correctness, which only adds a small overhead compared to the actual communication since it is just a scalar. After checking the correct usage of the primitives, users may also easily turn off this feature to enable more efficient communication.

Tensor fusion is another technique to boost the performance, which batches multiple small tensor communication requests into one and sends the fused tensor instead \cite{sergeev2018horovod}. To achieve that, it executes the following three steps: 1) copies multiple tensors into a continuous memory buffer, 2) communicates the tensor in the buffer, 3) copies the values in the buffer back to the original tensors. BlueFog supports tensor fusions for {\ttfamily allreduce}, {\ttfamily neighbor\_allreduce}, and {\ttfamily hierarchical\_neighbor\_allreduce}. The detailed implementation of tensor fusion and usage among them are different. Recall that the main motivation of tensor fusion is that ring-allreduce achieves the optimality under homogeneous bandwidth when the message is sufficient larger. Otherwise, the long latency of ring-allreduce can degrade the performance. Sacrificing a little waiting and copy time is worth for reducing the latency of ring-allreduce. However, neighborhood communication is always $O(1)$ delay. This nature of difference indicates a smaller buffer size for {\ttfamily neighbor\_allreduce} to achieve best performance. Moreover, the tensor fusion under dynamic neighbor topology are more complicated since the fused information are entangled with varying neighbor information.

\subsection{Putting Everything Together} 
Let's take a standard road path of one BlueFog communication as an example to understand how low-level collective communication APIs of BlueFog work step-by-step when all the components above are put together. In the beginning, at application layer, users may call BlueFog collective communication functions, like {\ttfamily allreduce}, {\ttfamily neighbor\_allreduce} together with topology associated information, necessary for dynamic usage. After that, the communication request and input tensor are transferred into system layer and then pushed into the shared queue waiting for the communication thread to handle it. The communication thread dequeues the request and sends the request to the negotiation service that holds at rank 0. The negotiation daemon has lots of responsibilities such as sorting the order of execution, checking the topology information, fusing many small requests into one, etc. As soon as the negotiation service collects requests from all ranks, it broadcasts a signal to all ranks that they are ready to execute the actual communication through MPI or NCCL depending on the tensor types, requests, etc. In the end, a callback function passed through the request is called to inform the application layer that the communication is finished.

%\vspace{-2mm}

% It is worth to remark that using AWC style algorithm, {\it hierarchical\_neighbor\_allreduce} can no longer form a super node as we original design idea. Because after the local gradient step, the nodes within the same machine will diverge again. Based on our experiment, this divergence may deteriorate the performance and become even worse when the task is training deep neural network. 

\section{Experimental Evaluation} \label{sec-evaluation}
% In our evaluation, we study the following questions:

% 1. How well is our neighbor allreduce compared with allreduce in terms of communication time under different topologies on CPUs and GPUs?

% 2. How does BlueFog provide a flexible usage to solve large-scale decentralized optimization problems?

% 3. In terms of large-scale deep neural network training task, how does BlueFog out-perform
% other state-of-the-art systems like Horovod in training speed and evaluation accuracy? 

All experiments in this section  are run on Amazon Web Services (AWS)\cite{aws}. Unless it is stated in other places, we use m4.4xlarge instance for CPU usage and p3.16xlarge instance for GPU usage \cite{AWS-p3}. Each m4.4xlarge server has 16 CPU cores and each p3.16xlarge server contains 8 NVIDIA V100 GPUs.

\subsection{Microbenchmarks}
%JL: Low level API latency, throughput\\

In this section, we compare the communication time between {\ttfamily neighbor\_allreduce} and {\ttfamily allreduce}. We carry out two experiments with CPUs and GPUs on AWS. In each experiment, we utilize the three different communication approaches to process the synthetic data of 1 megabytes (MB) for CPUs and 10 MB for GPUs as GPUs typically handle bigger computing capacity.

\begin{figure}[t]
    \centering
    \includegraphics[width=0.37\textwidth]{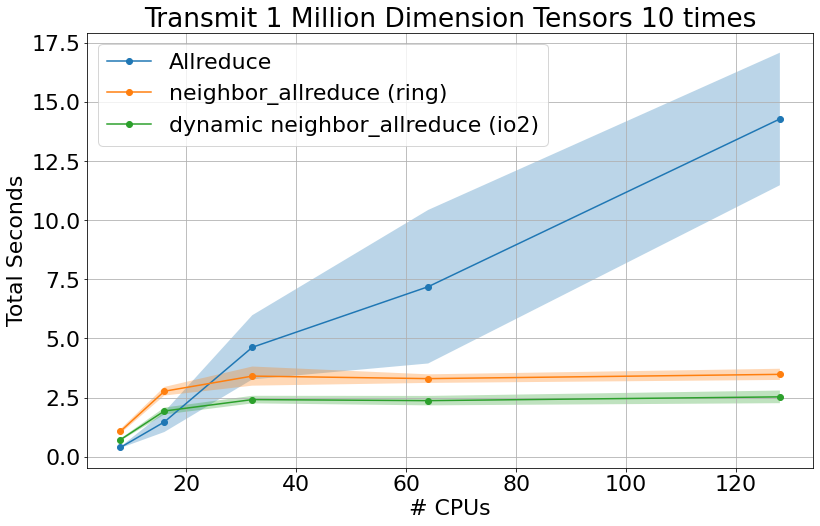}\vspace{-0.2mm}
    \includegraphics[width=0.37\textwidth]{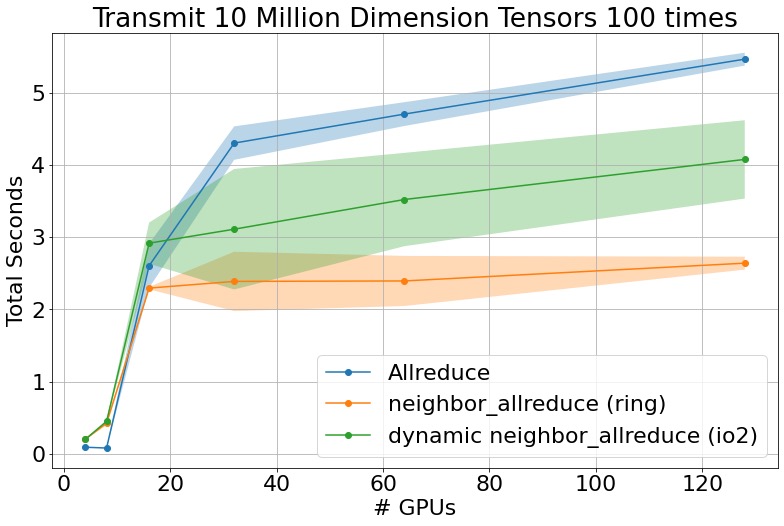} \vspace{-2mm}
    \caption{\small The execution time of allreduce, neighbor allreduce, and dynamic neighbor allreduce based on CPUs (left) and GPUs (right). %The execution time corresponds to the total time to communicate data 10 times.
    We perform the execution 10 times to calculate the average execution time represented by the solid points while the shaded areas represent 90\% confidence interval.}
    \label{fig:micro_benchamrk}
\end{figure}

% \begin{figure}[t]
%     \centering
%     \caption{ The execution time of allreduce, neighbor allreduce, and dynamic neighbor allreduce based on GPUs. %The execution time corresponds to the total time to communicate data 100 times.
%     We perform the execution 10 times to calculate the average execution time represented by the solid points while the shaded areas represent 90\% confidence interval.}
%     \label{fig:gpu_micro_benchamrk}
% \end{figure}

To fairly compare different neighbor allreduce methods, we select neighbor allreduce on the ring topology and dynamic neighbor allreduce on the inner-outer exponential-2 graph, supported by BlueFog, so that the data size for transfer in each iteration matches. As we can see from Fig. \ref{fig:micro_benchamrk}, the proposed neighbor allreduce primitives in BlueFog takes much less time for communication than allreduce on both CPUs and GPUs, especially with more computation cores. In addition, the time consumption for neighbor allreduce increases much slowly compared to allreduce as the number of cores increases, indicating better scalability of the neighbor communication methods.
%suggested in Section \ref{sec-dec-opt-impl}.

Note that the communication speed on GPU significantly drops from 8 GPUs to 16 GPUs for all the three methods. Remember that a single p3.16xlarge instance only contains 8 GPUs. We conjecture that the high-speed NVLink significantly boosts the communication efficiency within the local machine, while the communication across multiple machines becomes the bottleneck here, which is also supported in the later decentralized DNN benchmarks in Section \ref{sec:dnn_results} as well.

\subsection{Decentralized DNN training with BlueFog}
\label{sec:dnn_results}
% YBC + HBH: DNN: Throughput Communication VS Computation heavy

% Accuracy and Loss over time and epochs X ResNet + (BERT) 
\begin{figure*}[t]
    \centering
    \includegraphics[width=0.32\textwidth]{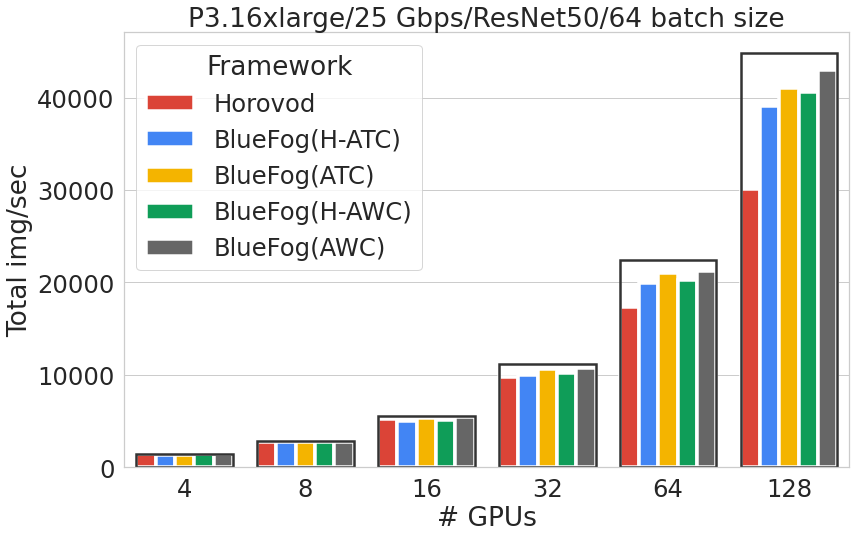}
    \includegraphics[width=0.32\textwidth]{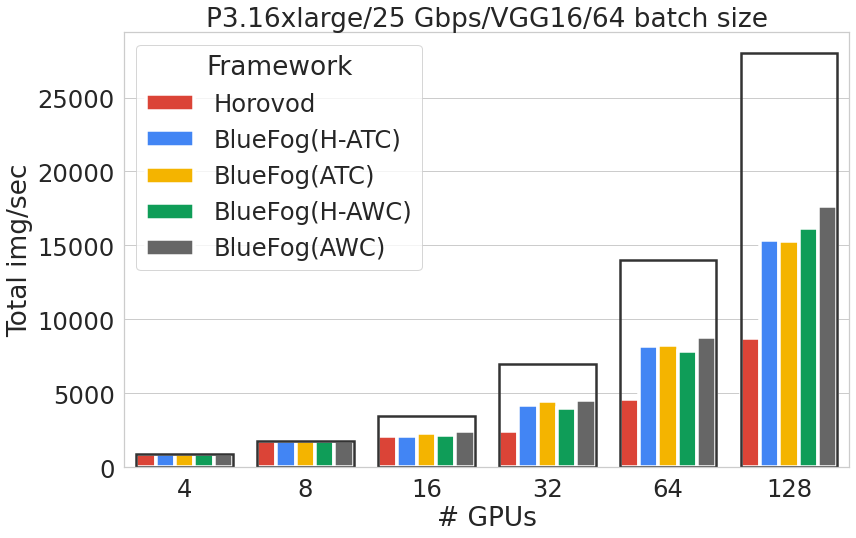}
    \includegraphics[width=0.32\textwidth]{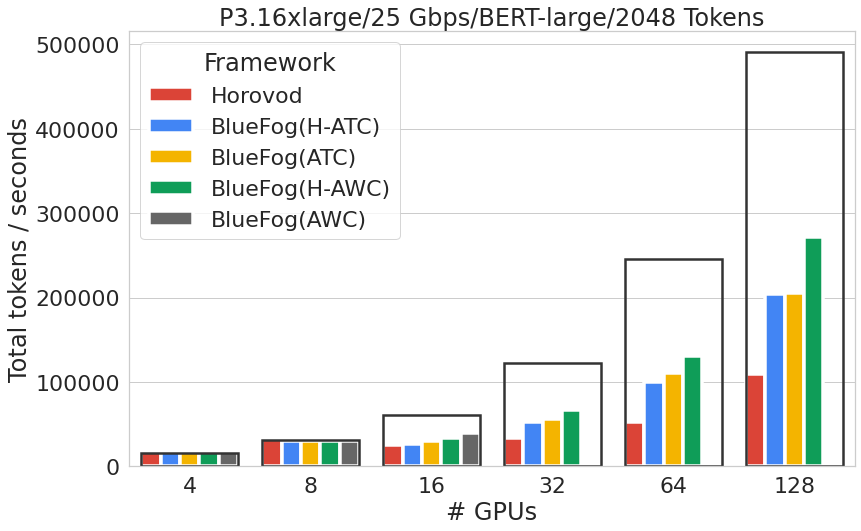}
    \vspace{-2mm}
    \caption{\small Throughput performance comparison over ResNet-50, VGG-16, and BERT-large models. 
    Label H-ATC represents the hierarchical neighbor allreduce with ATC-style over dynamic exponential 2 topologies. H-AWC is the corresponding AWC-style one. ATC and AWC represents the neighbor allreduce over dynamic inner and outer exponenetial 2 topology with ATC- and AWC- style algorithms.
    The batch-size or tokens are for one GPU. One machine has 8 GPUs. Hence, for 4 and 8 GPUs data-point, which corresponding to one machine, we use neighbor-allreduce result for hierarchical neighbor-allreduce's one.}
    \label{fig:dnn_benchamrk}
\end{figure*}
\begin{figure*}[htp]
    \centering
    \includegraphics[width=0.32\textwidth]{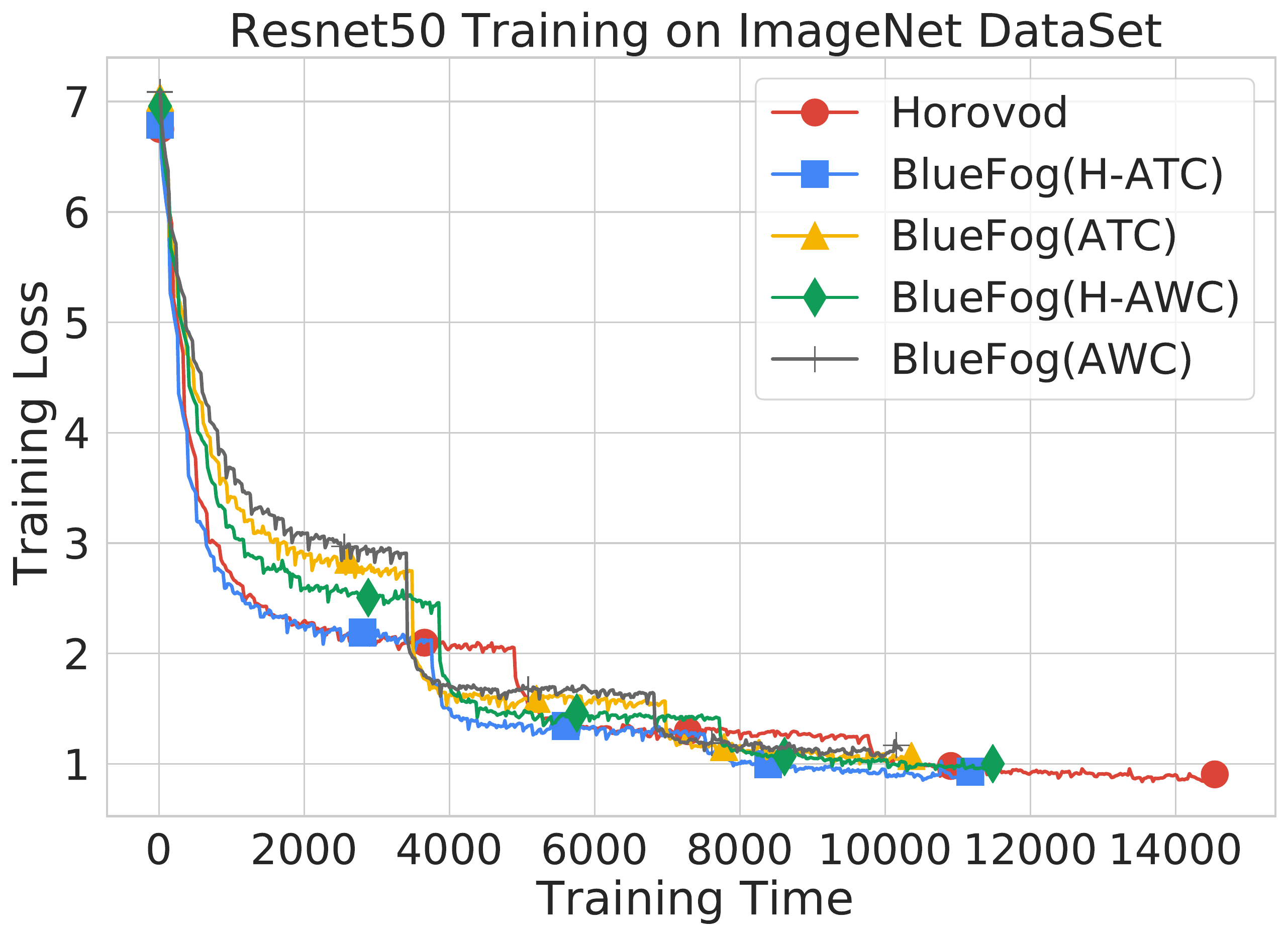}
    \includegraphics[width=0.32\textwidth]{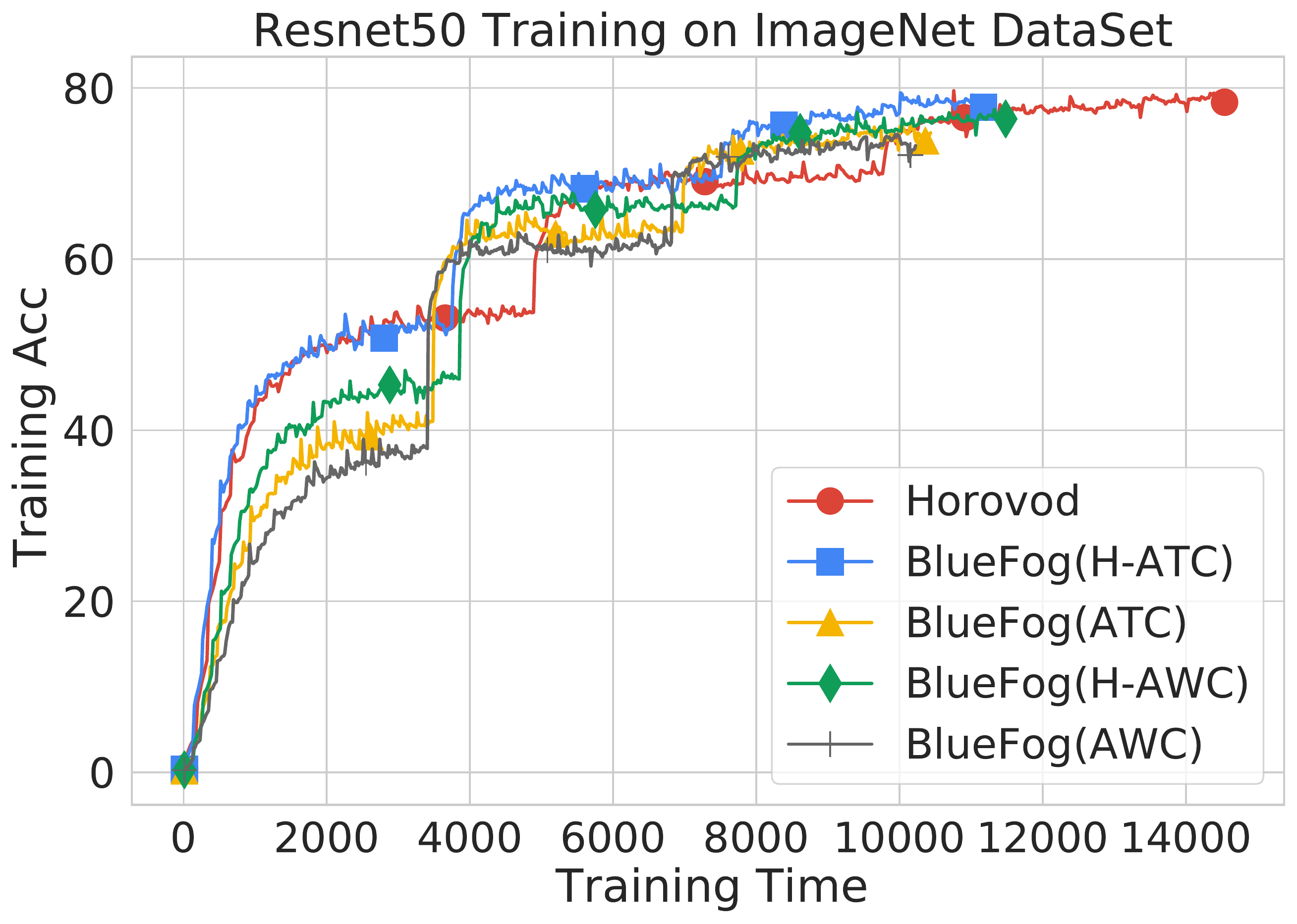}
    \includegraphics[width=0.32\textwidth]{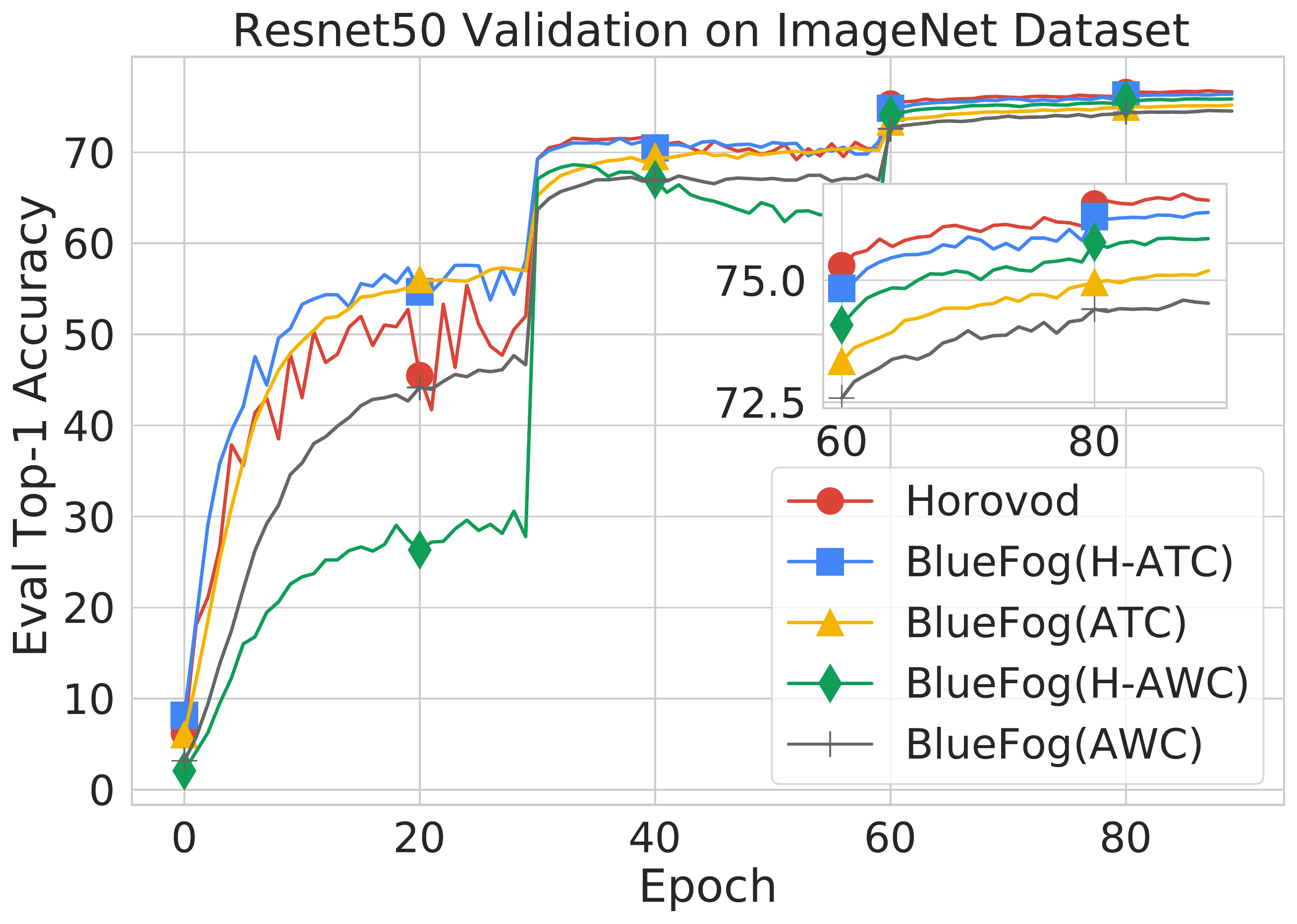}
    \caption{\small ImageNet training loss and accuracy in terms of wall clock time with ResNet50 and top-1 validation accuracy in terms of epochs. }
    \label{fig:imagenet_result}
    \vspace{-2mm}
\end{figure*}
% add a table [time and final accuracy]
% \begin{figure}[htp]
%     \centering
%     \includegraphics[width=0.45\textwidth]{Imagenet-resnet50-trainloss.pdf}
%     \caption{\small ImageNet training loss in terms of iterations and wall clock time with ResNet50 model. }
%      \label{fig:imagenet-train}
% \end{figure}

% \begin{figure}[htp]
%     \centering
%     \includegraphics[width=0.45\textwidth]{Imagenet-resnet50-acc-epoch.pdf}
%     \caption{\small Top-1 validation accuracy on the ImageNet with ResNet50 model. }
%     \label{fig:imagenet-eval-acc}
% \end{figure}

We demonstrate BlueFog’s performance on decentralized DNN training at different scales.
All experiments are done on AWS p3.16xlarge instances. Each instance has 8 Tesla V100 16GB GPUs connected through NVLink. Each pair of instances is connected through a network bandwidth of 25Gbps.

{\bf Throughput Benchmark.} To begin with, we test the throughput performance over three popular DNN models -- ResNet-50\cite{he2016deep}, VGG-16\cite{simonyan2014very}, and BERT-large\cite{devlin2018bert}. First two are both very classical models for image classification and BERT-large is one of the most popular models for natural language processing. The results are shown in \liu{ Fig.}~\ref{fig:dnn_benchamrk}. We use Horovod+NCCL as the baseline and run 4 BlueFog supported algorithms, of which are the AWC-style and ATC-style algorithms using {\ttfamily neighbor\_allreduce} and {\ttfamily hierarhical\_neighbor\_allreduce} APIs with a dynamic exponential-2 topology respectively \cite{ying2021exponential}. For shortage, we will call them BlueFog(AWC), BlueFog(ATC), BlueFog(H-AWC), and BlueFog(H-ATC).

In term of throughput, the results are quite consistent. No matter which algorithm BlueFog uses, it is always faster than allreduce-style framework, especially when the network size becomes larger. When training with 128 GPUs, the speedup of BlueFog over all-reduce is 1.2x to 1.8x. This is not surprising since the theoretical communication cost and the results in micro-benchmark have already shown that the neighborhood communication is much cheaper than the global allreduce communication. From 4 GPUs to 128 GPUs, the speedup becomes larger. It is reasonable to expect that the advantage of using BlueFog will be even larger with more GPUs.

We do observe that the benchmark we report is relative low in terms of linear scaling up. In the ResNet-50 experiment, BlueFog can reach over 95\% scaling efficiency on 128 GPUs, while in VGG and BERT-large experiment, it only reachs around 50-60\% efficiency. It is mainly because the experiment environment is 25Gbps without RDMA, which can become the bottleneck of linear scaling up especially for the computation intensive model like BERT-large. In \liu{ Fig.}~\ref{fig:dnn_benchamrk},  we can observe the scaling efficiency \liu{dramatically dropped} from 8 to 16 GPUs, which \liu{corresponds} to \liu{increasing} one AWS p3.16xlarge instance to two instances. Within a single instance, the communication between processes is through the high-speed NVLink, which is way faster than inter-machine network. 
Moreover, note that RestNet-50 has around 23 million parameters, VGG-16 has 138 million parameters, and BERT-large has 345 million parameters. The scaling efficiency are also affected by the number of model parameters, relatively.

{\bf Learning Curves.} As Bluefog utilizes inexact averaging over different typologies, we further run a series of image classification experiments on ImageNet-1k\cite{deng2009imagenet} dataset to validate the performance and generalisation of our system. 
%The ImageNet-1k \cite{deng2009imagenet} dataset consists of 1,281,167 training images and 50,000 validation images in 1000 classes. 
We train ResNet-50 model following the training protocol of \cite{goyal2017accurate}. The Nesterov momentum SGD optimizer is used with a linear scaling learning rate strategy. 
% All experiments are conducted with 8 AWS p3.16xlarge instances and all data are stored inside the host instance.
The left and middle sub-figures in {Fig.}~\ref{fig:imagenet_result} \liu{show} the evolution of training loss and accuracy in terms of wall clock time. Compared to Horovod, our implementation of decentralized communication gains 1.3x - 1.43x speed-up with \liu{similar} convergence. The rightmost sub-figure in \liu{Fig.}~\ref{fig:imagenet_result} \liu{shows} the top-1 validation accuracy of the aforementioned decentralized training methods. 
% {\color{red} Compared to the Horovod baseline, the hierarchical version of ATC and AWC training methods reach similar accuracy, while the pure ATC and AWC undergo a slight accuracy deterioration. However, the gap could be mitigated with tuning of training schedule (e.g. more iterations) or adjusting the learning rate.}  
More convergence results of decentralized optimization algorithms applied on DNN training tasks have also been reported in \cite{assran2019stochastic, lian2018asynchronous,luo2020prague}, which should give similar results if implemented in BlueFog.

\begin{table}[]
    \caption{\small Comparison training time on 90 epochs ResNet-50 ImageNet and it corresponding validation accuracy between multiple algorithms.}
    \centering
    \begin{tabular}{@{\hspace{-0.1mm}}l@{\hspace{-0.1mm}}ccc}
    \hline
         Algorithms       &  Time(Sec.) & Val. Accuracy& Speed Up\\\hhline{====}
         Horovod & 14648.26  & 76.6\%   & 1.00x    \\\hline
         BlueFog(H-ATC) & 11255.90  & 76.4\%   & 1.30x    \\\hline
         BlueFog(ATC)     & 10437.62  & 75.2\%   & 1.40x   \\\hline
         BlueFog(H-AWC)   & 11568.88  & 75.9\%   & 1.26x     \\\hline
         BlueFog(AWC)     & 10228.92  & 74.5\%   & 1.43x     \\\hline
    \end{tabular}
    % \vspace{-3mm}
    \label{tab:resnet50}
    % \vspace{-4mm}
\end{table}

\textbf{More models and algorithms.} We further examine the BlueFog over multiple models -- ResNet \cite{he2016deep}, MobileNetv2 \cite{sandler2018mobilenetv2}, and EfficientNet \cite{tan2019efficientnet}, and multiple decentralized algorithms -- the vanilla DmSGD \cite{assran2019stochastic}, DmSGD \cite{yu2019linear}, and QG-DmSGD \cite{lin2021quasi}. The task is the same as the ImageNet classification described before. Table \ref{tb:8*8} lists the top-1 validation accuracy comparison across all the models and algorithms. We also list the performance of the parallel SGD using global averaging as the baseline. For each model and algorithm, we examine it over two topology settings, one is the static exponential topology and the other is the dynamic exponential topology. In the latter topology,  each process only picks one neighbor at each iteration \cite{ying2021exponential}. The table \ref{tb:8*8} shows that the dynamic topology can further reduce the communication cost without any noticeable performance degrade, which is one main reason that BlueFog is designed to support dynamic topologies. The complete freedom of communication control encourages users to develop more efficient decentralized algorithms to train DNNs.

\begin{table}[htp]

\begin{center}
\vskip -2mm
\caption{\small \color{black} Top-1 validation accuracy and wall-clock time (in hours) comparison with different models and algorithms on ImageNet dataset over static/dynamic exponential topology (8x8 GPUs)\cite{ying2021exponential}. }
\vskip -1mm

\label{tb:8*8}
\begin{small}
\begin{sc}
% \centering
\setlength{\tabcolsep}{1.2mm}{
\begin{tabular}{ccccccc}
\toprule
    model &  \multicolumn{2}{c}{ResNet-50} & \multicolumn{2}{c}{MobileNet-v2} & \multicolumn{2}{c}{EfficientNet} \\
    Topology & static & dynamic & static & dynamic  & static & dynamic \\
    \midrule
    Parallel SGD &  76.21 (7.0) & - &    70.12 (5.8) & -  & 77.63 (9.0) & -  \\ 
    vanilla DmSGD & 76.14 (6.6) & 76.06 (5.5) & 69.98 (5.6) & 69.81 (4.6)  & 77.62 (8.4) & 77.48 (6.9)  \\
    DmSGD & 76.50 (6.9) & 76.52(5.7)  & 69.62 (5.7) & 69.98 (4.8) & 77.44 (8.7) & 77.51 (7.1)  \\
    QG-DmSGD & 76.43 (6.6) & 76.35(5.6)  & 69.83 (5.6) & 69.81 (4.6)  & 77.60 (8.4) & 77.72 (6.9)  \\
\bottomrule
\end{tabular}}
\end{sc}
\end{small}
\end{center}

\vskip -5mm
\end{table}

\section{Conclusion}\label{sec-conclusion}
{
In this paper, we present BlueFog, an open-source library for efficient and high-performance implementation of decentralized algorithms in optimization and deep learning. Through a unified abstraction of different decentralized communication operations, BlueFog provides simple and consistent communication primitives to support diverse decentralized algorithms. BlueFog can be used with PyTorch to train deep neural networks. The system design philosophy and detailed implementation of BlueFog are carefully presented to demonstrate the superior performance. The usage of BlueFog is illustrated with various application examples in optimization and signal processing. Deep learning experiments support that BlueFog outperforms Horovod, a state-of-the-art distributed training framework based on Ring-Allreduce. 

% , which fills the gap between the decentralized optimization algorithms and deep learning training.
% A unified  API {\it neighbor\_allreduce} and its hierarchical version are introduced, supporting neighborhood communication over arbitrary static and dynamic topologies. Hence, BlueFog enables fast implementation of a variety of decentralized optimization algorithms and efficient distributed DNN training. 
% %In addition, we also prove that dynamic exponential 2 topology can achieve the .
% Leveraging the communication efficiency of dynamic topology strategy, we have shown that BlueFog outperforms the state-of-the-art distributed training framework in many popular deep learning models. 
}

\section{Acknowledgement}
The authors would like to thank Dr. Ji Liu from Baidu Inc. for his contribution to BlueFog and the discussion on the early version of this manuscript, and Edward Nguyen from University California, Los Angeles for contributing to the tutorials on BlueFog.

%-------------------------------------------------------------------------------
\bibliographystyle{IEEEbib}
\bibliography{references}

\newpage
% {\color{white} blank }
% \newpage 
\appendix

\subsection{Application example: Exact-Diffusion algorithm}\label{app-exact-diffusion}
This section still considers the decentralized linear regression problem \eqref{dlr}. We will show how to implement Exact-Diffusion \cite{yuan2017exact1,li2019decentralized}, a decentralized algorithm that can correct the bias suffered by decentralized gradient descent, over static topology. 

\vspace{1mm}
\noindent \textbf{Exact-Diffusion.} Exact-Diffusion with static graph topology will iterate as follows: 
\begin{align}
\psi_i^{(k)} &= x_i^{(k)} - \gamma A_i^T(A_i x_i^{(k)} - b_i)  \hspace{2cm} \mbox{(local update)} \label{ed-1}\\
\phi_i^{(k)} &= \psi_i^{(k)} + x_i^{(k)} - \psi_i^{(k-1)}  \hspace{2.65cm} \mbox{(bias correction)} \label{ed-2}\\
x_i^{(k+1)} &= w_{ii} \phi_i^{(k)}+ \sum_{j \in \cN(i)} w_{ij} \phi_j^{(k)}  \hspace{2.15cm} \mbox{(partial averaging)} \label{ed-3}
\end{align}
where $\cN(i)$ is the in-coming neighbors of node $i$. 

\vspace{1mm}
\noindent \textbf{Code.} We set the topology as the static ring graph in the following Exact-Diffusion  implementation. The code snippet using BlueFog is shown in Listing \ref{lst.ED-static}. The complete code can be referred to BlueFog online tutorial\footnote{\url{https://github.com/Bluefog-Lib/bluefog-tutorial/tree/master/Section\%204}}. 

\vspace{2mm} 
\begin{lstlisting}[language=python, numbers=left, captionpos=b, label={lst.ED-static},
caption={\small BlueFog implmentation of the Exact-Diffusion algorithm to solve decentralized linear regression.}]
import bluefog.torch as bf
from bluefog.common import topology_util
bf.init()  # Initialize the BlueFog

# Set topology as static ring graph.
G = topology_util.RingGraph(bf.size())
bf.set_topology(G)

# ED implementation
for ite in range(maxite):
    grad_local = A.t().mm(A.mm(x) - b)  # compute local grad
    psi = x - gamma * grad_local        # local update
    phi = psi + x - pre_psi             # bias correction
    x = bf.neighbor_allreduce(phi)      # partial averaging
    pre_psi = psi.clone()             
\end{lstlisting}

\subsection{Application example: Push-sum gradient tracking algorithm}\label{app-push-sum}
This section considers the decentralized linear regression problem \eqref{dlr}. We will show how to implement push-sum gradient tracking \cite{nedic2017achieving} in BlueFog, a decentralized algorithm based on the push-style communication, over time-varying topologies. 

\vspace{1mm}
\noindent \textbf{Push-sum gradient tracking.} The push-sum gradient tracking over time-varying topology will iterate as follows: 
\begin{align}
u_i^{(k+1)} &= w^{(k)}_{ii} (u_i^{(k)} - \gamma y_i^{(k)} ) + \sum_{j \in \cN(i)} w^{(k)}_{ij} (u_j^{(k)} - \gamma y_j^{(k)})\\
v_i^{(k+1)}  &= w^{(k)}_{ii} v_i^{(k)}  + \sum_{j \in \cN(i)} w^{(k)}_{ij} v_j^{(k)}\\
x_i^{(k+1)} &= u_i^{(k+1)}/v_i^{(k+1)} \\
g_i^{(k+1)} &=  A_i^T(A_i x_i^{(k+1)} - b_i)  \\
y_i^{(k+1)} &= w^{(k)}_{ii} (y_i^{(k)} + g_i^{(k+1)}  - g_i^{(k)} )  + \sum_{j \in \cN(i)} w^{(k)}_{ij} \big(y_j^{(k)} + g_j^{(k+1)}  - g_j^{(k)} \big)
\end{align}
where $\cN(i)$ is the in-coming neighbors of node $i$, and $y_i^{(0)} = g_i^{(0)}$

\vspace{1mm}
\noindent \textbf{Code.} We set the topology as the one-peer variant of grid topology. The code snippet using BlueFog is shown in Listing \ref{lst.GT-varying}. The complete code can be referred to BlueFog online tutorial\footnote{\url{https://github.com/Bluefog-Lib/bluefog-tutorial/tree/master/Section\%205}}. 

\vspace{2mm} 
\begin{lstlisting}[language=python, numbers=left, captionpos=b, label={lst.GT-varying},
caption={\small BlueFog implmentation of the push-sum Gradient-Tracking algorithm over time-varying topologies.}]
import bluefog.torch as bf
from bluefog.common import topology_util
bf.init()  # Initialize the BlueFog

def Gradient_tracking_one_step(
    x, y, u, v, prev_grad_local, dynamic_neighbor_gen, x_opt, A, b, gamma=1e-2):

    # Get the update weights
    dst_neighbors, src_neighbors = next(dynamic_neighbor_gen)
    src_weights = {r: 1 for r in src_neighbors}
    self_weight = 1 / (len(dst_neighbors) + 1)
    dst_weights = {r: self_weight for r in dst_neighbors}

    # u update step: u_{k+1} = W^k (u_k - \alpha*y_k)
    w = u - alpha * y
    u_new = bf.neighbor_allreduce(w, self_weight, src_weights, dst_weights)
    
    # v update:  v_{k+1} = W^k v_{k}  ---   (Correction weights)
    v_new = bf.neighbor_allreduce(v, self_weight, src_weights, dst_weights)
    
    # x update: x_{k+1} = u_{k+1} / v_{k+1} (element-wise division)
    x_new = u_new / v_new
    
    # y update step: y_{k+1} = W^k (y_k + \grad f(x_{k+1}) - \grad f(x_{k}))
    grad_local = A.t().mm(A.mm(x_new) - b)
    q = y + grad_local - prev_grad_local
    y_new = bf.neighbor_allreduce(q, self_weight, src_weights, dst_weights)
    
    return x_new, y_new, u_new, v_new, grad_local
   
# Set up grid topology
bf.set_topology(topology_util.MeshGrid2DGraph(bf.size()))
# Set up one-peer variant of the grid topology
dynamic_neighbor_gen = topology_util.GetDynamicOnePeerSendRecvRanks(bf.load_topology(), bf.rank())

# Algorithm starts
for ite in range(maxite):
    x, y, u, v, prev_grad_local = Gradient_tracking_one_step(
        x, y, u, v, prev_grad_local, dynamic_neighbor_gen, x_opt, A, b, gamma=2e-2)  
\end{lstlisting}

\end{document}